\documentclass[12pt]{article}
\usepackage{cite}
\usepackage{graphicx,bbm,pstricks,soul}
\usepackage{pifont}
\usepackage{bbm,algorithmic,mdframed,placeins,multirow,booktabs,subfigure}
\usepackage[ruled,vlined,linesnumbered]{algorithm2e}
\usepackage{tikz,hhline}
\usepackage{tabularx}
\newcolumntype{L}[1]{>{\raggedright\let\newline\\\arraybackslash\hspace{0pt}}m{#1}}
\newcolumntype{C}[1]{>{\centering\let\newline\\\arraybackslash\hspace{0pt}}m{#1}}
\newcolumntype{R}[1]{>{\raggedleft\let\newline\\arraybackslash\hspace{0pt}}m{#1}}

\usepackage{amsmath,amsfonts,amsbsy,amssymb}

\newcommand{\LRARR}[4]{{\mbox{ \raise 0.4 mm \hbox{$#1$}}} \;
  \mathop{\stackrel{\displaystyle\longrightarrow}\longleftarrow}^{#3}_{#4}
  \; {\mbox{\raise 0.4 mm\hbox{$#2$}}}}

\newcommand{\picturesAB}[6]{
\centerline{
\hskip #4
\raise #3 \hbox{\raise 0.9mm \hbox{(a)}}
\hskip #5
\epsfig{file=#1,height=#3}
\hskip #6
\raise #3 \hbox{\raise 0.9mm \hbox{(b)}}
\hskip #5
\epsfig{file=#2,height=#3}
}}
\newcommand{\picturesCD}[6]{
\centerline{
\hskip #4
\raise #3 \hbox{\raise 0.9mm \hbox{(c)}}
\hskip #5
\epsfig{file=#1,height=#3}
\hskip #6
\raise #3 \hbox{\raise 0.9mm \hbox{(d)}}
\hskip #5
\epsfig{file=#2,height=#3}
}}

\makeatletter  
\newcommand{\xleftrightarrows}[2][]{\mathrel{%
 \raise.40ex\hbox{$  
       \ext@arrow 3095\leftarrowfill@{\phantom{#1}}{#2}$}%
 \setbox0=\hbox{$\ext@arrow 0359\rightarrowfill@{#1}{\phantom{#2}}$}%
 \kern-\wd0 \lower.4ex\box0}}  
 
\newcommand{\xrightleftarrows}[2][]{\mathrel{%
 \raise.40ex\hbox{$\ext@arrow 3095\rightarrowfill@{\phantom{#1}}{#2}$}%
 \setbox0=\hbox{$\ext@arrow 0359\leftarrowfill@{#1}{\phantom{#2}}$}%
 \kern-\wd0 \lower.4ex\box0}}  
 
\def\leftrightarrowfill@{%
 \arrowfill@\leftarrow\relbar\rightarrow%
 }
\newcommand*{\centerfloat}{%
  \parindent \z@
  \leftskip \z@ \@plus 1fil \@minus \textwidth
  \rightskip\leftskip
  \parfillskip \z@skip}
\makeatother
\makeatother

\newtheorem{dfn}{Definition}[section]

\newcommand{\ltri}{%
\,\resizebox{!}{0.25\baselineskip}{%
\begin{tikzpicture}%
\draw[line width=1mm](0,0) -- (0,2) -- (2,0)  -- (0,0);
\end{tikzpicture}%
}\xspace%
}%

\newcommand{\smallltri}{%
\,\resizebox{!}{0.15\baselineskip}{%
\begin{tikzpicture}%
\draw[line width=1mm](0,0) -- (0,2) -- (2,0)  -- (0,0);
\end{tikzpicture}%
}\xspace%
}%

\author{Simon L. Cotter\thanks{School of
    Mathematics, University of Manchester, Manchester, UK. e:
    simon.cotter@manchester.ac.uk. SLC is grateful for EPSRC First
    grant award EP/L023393/1. SLC would like to thank the Isaac Newton
    Institute for Mathematical Sciences for support and hospitality
    during the programme ``Uncertainty quantification for complex systems: theory and methodologies'' when work on this paper was undertaken. This work was supported by:
EPSRC grant number EP/K032208/1.} \and Ioannis
G. Kevrekidis\thanks{Chemical and Biomolecular Engineering \& Applied
  Mathematics and Statistics, Johns Hopkins University, Baltimore, MD
  21218. The work of IGK was partially supported by the US National
  Science Foundation, by the US AFOSR, and by DARPA.} \and Paul
  Russell\thanks{School of
    Mathematics, University of Manchester, Manchester, UK.}}

\title{Transport map accelerated adaptive importance sampling, and application to inverse problems arising from
  multiscale stochastic reaction networks}
\begin{document}
\maketitle
\begin{abstract}
In many applications, Bayesian inverse problems can give rise to
probability distributions which contain complexities due to the
Hessian varying greatly across parameter space. This complexity often manifests itself as lower
dimensional manifolds on which the likelihood function is
invariant, or varies very little. This can be due to trying to infer
unobservable parameters, or due to sloppiness in the model which is
being used to describe the data. In such a situation, standard
sampling methods for characterising the posterior distribution, which
do not incorporate information about this structure, will be highly
inefficient. 

In this paper, we seek to develop an approach to
tackle this problem when using adaptive importance sampling methods,
by using optimal transport maps to simplify
posterior distributions which are concentrated on lower dimensional
manifolds. This approach is applicable to a whole range of
problems for which Monte Carlo Markov chain (MCMC) methods mix slowly.

We demonstrate the approach by considering inverse problems arising
from partially observed stochastic reaction networks. In particular,
we consider systems which exhibit multiscale behaviour, but for which
only the slow variables in the system are observable. We demonstrate
that certain multiscale approximations lead to more consistent
approximations of the posterior than others. The use of optimal
transport maps stabilises the ensemble transform adaptive importance
sampling (ETAIS) method, and allows for efficient sampling with
smaller ensemble sizes. This approach allows us to take
  advantage of the large increases of efficiency when using adaptive
  importance sampling methods for previously intractable Bayesian
  inverse problems with complex posterior structure. 
\end{abstract}

\section{Introduction}
The importance of Markov chain Monte Carlo (MCMC) methods is becoming
increasingly apparent, in a world replete with datasets which need to be
combined with complex dynamical systems in order for us to be able
make progress in a range of scientific fields. Different MCMC methods
have been designed with different challenges in mind, for example high dimensional
targets\cite{cotter2013mcmc}, or large data
sets\cite{bierkens2016zig}. New types of stochastic processes
have been considered, such as piecewise deterministic Markov processes
\cite{bouchard2018bouncy}.
Other methods are able to make very large moves informed by the
data\cite{teh2016consistency,duane1987hybrid}, and others are able to exploit
the geometry of the target\cite{girolami2011riemann}. Population Monte Carlo
(pMC) methods use efficient importance sampling proposals informed by an
ensemble of particles which have already learned about the structure
of the target\cite{cappe2012population,cappe2008adaptive,douc2007convergence,douc2007minimum, martino2017layered, cornuet2012adaptive, martino2015adaptive}. These methods are particularly useful in the context of
low dimensional target distributions which have complex\footnote{We
  use ``complex'' with the meaning ``complicated'' rather than
  containing imaginary numbers.} structure, for
which traditional Metropolis-Hastings methods tend to struggle, for
example when the target is multimodal or if the density is
concentrated on a lower-dimensional manifold. Moreover, they are an
excellent candidate for parallelisation, since the task of computing
the likelihood for each particle in the ensemble can be distributed
across a set of processors. This is of utmost importance in an age
where increases in computational power are most likely to come from
large-scale parallelisation as opposed to the large speed-ups in
individual processors that we saw throughout the 20th century.

In \cite{cotter2015parallel}, we explored ideas from adaptive MCMC
within adaptive importance samplers, in order to construct a method which could automatically adapt to optimise
efficiency. Alongside this, we explored the use of state-of-the-art
resamplers\cite{reich2013nonparametric}, and the multinomial transform
(MT), a greedy approximation of the ensemble transform method, in order to improve the quality of
the importance sampler proposal distribution.

Other approaches to challenging posterior structure have been
proposed, including the work of Marzouk and Parno\cite{parno2018transport}, who
demonstrated how invertible transport maps can be constructed which
map the target distribution close to an easily-explored reference
density, such as a $d$-dimensional Gaussian. Since the map is
invertible, Gaussian proposal densities on the reference space can be
mapped to highly informed moves on the parameter space. A related idea
is the use of the anisotropic diffusion maps\cite{singer2008non}, in
which a map which describes the manifold is constructed using
Principle Component Analysis. These ideas have also been applied to
stochastic reaction
networks\cite{singer2009detecting,dsilva2016data}. In
\cite{chiavazzo2017intrinsic}, these maps were used to propagate
data-mining into unexplored regions which were predicted to be close
to the lower-dimensional manifold. The idea is similar to that of the
Lamperti or Girsanov transformations for SDEs\cite{flandrin2003stationarity,gardiner2009stochastic},  the aim being to map complex
processes onto a standard one which is well understood, through a
change of measure.

There are many applications in which complex posterior structures
often arise. One family of such problems is those for which the
forward model is insensitive to changes in certain directions of
parameter space, sometimes called
\emph{sloppiness}\cite{gutenkunst2007universally,apgar2010sloppy}. This
sloppiness in the forward model appears in many applications in
science and engineering\cite{constantine2014active}. Where one is
aiming to solve an inverse problem based on such a model, this can
lead to the posterior distribution being highly concentrated on lower
dimensional manifolds, which represent the directions of
sloppiness. This is particularly prevalent in partially observable
systems, for example multiscale systems such
as stochastic reaction networks, where only slowly changing quantities
can be reliably observed.

In this paper, we aim to incorporate transport maps into an ensemble
importance sampling setting, in order to develop stable algorithms which are
capable of sampling from probability distributions with complex
structures. These maps are constructed using an importance sample of
the posterior distribution, and are used to simplify the posterior
distribution, mapping them as close as possible to a reference
Gaussian measure. Through this map, straightforward Gaussian proposal
densities are transformed into complex densities, which match as well
as possible the target density. Less well-informed proposal kernels means that complex
structures in the density can only be well represented with a large
increase in the ensemble size, increasing computational cost. If
computational resources do not allow for such an increase, this can
lead to stability issues in the algorithm, due to the proposal
distribution in the importance sampler not representing well the
target density, leading to high variance in the weights, causing slow
convergence. Better informed proposal densities
lead to more stable and faster converging importance sampling
schemes, and we will demonstrate this through examples
arising from Bayesian inverse problems on multiscale stochastic
reaction networks.

In a perfect scenario, the numerical
  approximation of the optimal transport map would capture all of the
  posterior structure, and independent samples could be drawn by
  sampling from the reference measure. However in many cases the pushforward of the posterior measure can still exhibit complex structure
  which means that efficient adaptive importance sampling can be
  preferential to this independent sampling approach.


In Section \ref{sec:ETAIS}, we will briefly reintroduce the ensemble
transform adaptive importance sampling (ETAIS) method. In Section \ref{sec:map} we show how an appropriate transport map can
be constructed from importance samples which maps the posterior close
to a reference Gaussian measure. In Section \ref{sec:TETAIS} we show
how such a map can be incorporated into a sophisticated parallel MCMC
infrastructure in order to accelerate mixing, and improve stability. In Section
\ref{sec:conv} we seek to show the advantages of this approach through
the analysis of a challenging test problem. In Section
\ref{sec:multi} we consider how likelihoods can be approximated using
 multiscale methodologies in order to carry out inference for
multiscale and/or partially observed stochastic reaction networks. In
Section \ref{sec:num} we present some numerical examples, which serve
to demonstrate the increased efficiency of the described sampling
methodologies, as well as investigating the posterior approximations
discussed in the previous section. We conclude with a discussion in
Section \ref{sec:conc}.

\section{Ensemble Transform Adaptive Importance Sampling}\label{sec:ETAIS}
Ensemble transform adaptive importance sampling
(ETAIS)\cite{cotter2015parallel} is an adaptive importance sampling framework
which uses an ensemble of particles and state-of-the-art resampling
methods to construct a mixture proposal distribution which closely
matches the target distribution. 

In importance sampling, we aim to sample from a target $\pi$, by
sampling from a chosen proposal distribution $\chi$. Each sample $\theta^{(k)} \sim \chi$ is then weighted by
$w_k = \frac{\pi(\theta^{(k)})}{\chi(\theta^{(k)})}$, to take account of the bias of
sampling from a different distribution to $\pi$. Monte Carlo estimates
using a sample of size $N$
of a function $f$ with respect to $\pi$ can then be made through the
formula
\begin{equation} \mathbb{E}_\pi(f) \approx \frac{1}{\bar{w}} \sum_{k=1}^N
  w_kf(\theta^{(k)}),\end{equation}
where $\bar{w} = \sum_{k=1}^N w_k$ is the total weight.
This method works well when $\pi$ and $\chi$ are close, but can be
excruciatingly slow when they are not. The idea behind adaptive
importance sampling methods is
to construct a good proposal distribution, either from the entire
history of the algorithm up to the current point, or to use the
current state of a whole ensemble of $M$ particles in the system.  {A nice illustrative figure of how adaptive
  importance sampling algorithms work is given in \cite{bugallo2017adaptive}.}

In ETAIS, the proposal distribution $\chi$ is chosen to be the equally
weighted mixture of a set of proposal kernels. If $\theta^{(k)} = [\theta_1^{(k)},
\theta_2^{(k)}, \ldots, \theta_M^{(k)}]^\top$ is the current state of the
ensemble, and we wish to use a mixture of the proposal density $q(\cdot ;
\cdot, \beta)$ then 
\begin{equation}\chi^{(k)} = \frac{1}{M} \sum_{i=1}^M q(\cdot ; \theta_i^{(k)},
\beta).\end{equation}
Usually the kernel $q(\cdot ;
x,\beta)$ is centred at, or has mean equal to $x$, and whose variance
scales with $\beta$. Good values for the
algorithmic parameter(s) $\beta$ can be
found by optimising for the effective sample size of the importance
sample that is produced (see \cite{cotter2015parallel, russ2017parallel} for more details).

If the ensemble is large enough, and the chain has entered
probabilistic stationarity, then the current state of the ensemble is
a good rough discrete approximation of the target density, and in turn
$\chi^{(k)}$ is close enough to $\pi$ to produce an efficient importance
sample $\{(\hat{\theta}^{(k)},w^{(k)})\}$,  where $\theta^{(k)} =
[\theta_1^{(k)}, \ldots, \theta_M^{(k)}]^\top$ and $w^{(k)} =
[w_1^{(k)}, \ldots, w_M^{(k)}]^\top$. It can be advantageous to use stratified sampling of the
mixture in order to
ensure that the sample made is as representative as possible of the
target density, i.e.
\begin{equation} \hat{\theta}_i^{(k)} \sim q(\cdot ; \theta_i^{(k)},\beta)\end{equation}
for each $i = 1,2,\ldots,M$.  We now have a weighted sample, and it
would be inadvisable to use an equally weighted mixture of proposal
distributions from each of these points. Therefore, before starting
the next iteration, the importance sample
$\{(\theta^{(k)},w^{(k)})\}$ is resampled to produce an equally
weighted sample, ready for the next iteration of the algorithm. In
ETAIS, a state-of-the-art resampler is used, which uses optimal
transport methods to find the equally weighted discrete sample which
best represents the statistics of the importance
sample\cite{reich2013nonparametric}. For larger ensemble sizes $M$ this can become
expensive, in which case a greedy approximation of this algorithm, the
Multinomial Transformation (MT) can be
implemented\cite{cotter2015parallel}. The output of the resampler is then denoted
$\theta^{(k+1)}$ and the algorithm is ready for the next
iteration. The ETAIS is summarised in Algorithm \ref{alg:ETAIS}, where
the importance samples $\{(\hat{\theta}^{(k)},w^{(k)})\}$ are stored as
the output.

\begin{table}[!h]
\centering
\begin{algorithm}[H]
\DontPrintSemicolon
\BlankLine
	Initialise $\theta^{(1)} \sim \pi_0$.\;
	\For{$k=1,\ldots,N$}{
		Sample $\hat{\theta}_i^{(k)} \sim q(\cdot; \theta_i^{(k)}, \beta)$, for $i = 1,\dots, M$.\label{algline:ETAIS_propose}\;
		Calculate $w^{(k)} = (w_1^{(k)}, \ldots, w_M^{(k)})^\top$, where
			\begin{equation}
				w^{(k)}_i = \frac{\pi(\hat{\theta}^{(k)}_i)}{\chi^{(k)}(\hat{\theta}^{(k)}_i; \theta^{(k)}, \beta)}.
			\end{equation}
		\label{algline:ETAIS_weights}

		Resample $\theta^{(k+1)} \leftarrow \| w^{(k)}\|_1^{-1}\sum_{j=1}^M w_j^{(k)}\delta_{\hat{\theta}_j^{(k)}}(\cdot)$.\label{algline:ETAIS_resample}\;
	}
	Output $\{(\hat{\theta}^{(n)},w^{(n)})\}_{n=1}^N$.
\caption{The ETAIS Algorithm.\label{alg:ETAIS}}
\end{algorithm}
\end{table}

{The choice of proposal kernel $q$ can affect the convergence of
  the algorithm. In particular, it is important that the tails of the
  proposal distribution are as fat as the target distribution. In this
  work we will primarily use Gaussian kernels, but we have previously
  made comparisons with more complex kernels, such as MALA and HMC\cite{cotter2015parallel,russ2017parallel}. There are minimal
  benefits from using these more complex kernels, which also come at
  additional cost.}

One problem with the ETAIS and other similar methods can become apparent if
the target density has highly curved manifolds on which the density is
concentrated. In this
case, unless the proposal densities $q$ are informed by this local
structure, the mixture distribution proposal may not well approximate
$\pi$ without a very large ensemble size $M$, which can become
prohibitively expensive. Some methods have been
proposed\cite{douc2007minimum}
which use samples local to each particle to inform local covariance
structure.

{In this paper, we investigate the use of transport maps which allows
us to learn the local correlation structures of the posterior in
currently explored regions. Using these transport maps, we can
stabilise adaptive importance sampling methods, and make these methods
more applicable to a wider range of more challenging inference problems.}

\section{Construction of transport maps in importance sampling} \label{sec:map}
In this Section, we describe the construction of transport maps which
allow for the simplification of complex posterior distributions in
order to allow for improved sampling, in particular for methods based
on importance sampling.
In~\cite{el2012bayesian} the transport map was introduced to provide a transformation from the prior
distribution to the posterior distribution, the idea being that one could draw a moderately sized
sample from the prior distribution and use this sample to approximate a map onto the target space.
Once this map was known to the desired accuracy, a larger sample from the prior could be used to
investigate the posterior distribution. This
methodology was adapted in~\cite{parno2018transport} to form a new proposal method for MH
algorithms. In this case, rather than transforming a sample from the prior into a sample from the target
distribution, the map transforms a sample from the posterior onto a reference space.
The reference density is chosen to allow efficient proposals using a simple proposal
distribution such as a Gaussian centred at the previous state. Proposed states can then be mapped back into a sample from the posterior by applying the inverse of the transport map.

Proposing new states in this way allows us to make large, well-informed
moves, even when sampling from probability distributions with complex structure.
If a very good such map can be found, it is also possible that the
pushforward of the posterior density through the map is close enough
to the reference Gaussian that we can efficiently
propose moves using a proposal distribution which is independent of
the current state, for example sampling from the reference Gaussian
itself, or a slightly more diffuse distribution.

In this Section we outline the methodology in
\cite{parno2018transport} for approximately coupling the target
measure with density
$\pi_{\theta}$, with the reference distribution with density $\pi_r$, and
show how the map can be constructed using a weighted sample
and hence how we can incorporate the map into importance sampling schemes.

\begin{dfn}[Transport Map $T$]
	A transport map $T$ is a smooth function $T\colon
        \mathcal{X}\rightarrow\mathbb{R}^d$ such that the {\it
          pullback} of the reference measure with density $\phi(\cdot)$,
	\begin{equation}\label{eq:pullback}
		\tilde{\pi}(\theta) = \phi(T(\theta))|J_T(\theta)|,
	\end{equation}
	is equal to the target density $\pi(\theta)$ for all $\theta \in \mathcal{X}$. The pullback is defined in terms of the determinant of the Jacobian of $T$,
	\begin{equation}
		|J_T(\theta)| = \text{det}\begin{bmatrix} \partial_{\theta_1} T_1(\theta) & \dots & \partial_{\theta_d} T_1(\theta) \\ \vdots & \ddots & \vdots \\ \partial_{\theta_1} T_d(\theta) & \dots & \partial_{\theta_d} T_d(\theta) \end{bmatrix}.
	\end{equation}
\end{dfn}


\begin{dfn}[Target and Reference Space]
	The transport map pushes a sample from a {\it target space} $\mathcal{X}$, that is a subset of $\mathbb{R}^d$ equipped with a target density $\pi_{\theta}$, onto a {\it reference space}, $R$, again a subset of $\mathbb{R}^d$ equipped with the reference density $\pi_r$.
\end{dfn}

Armed with such a map, independent samples can be made of the target
measure, using the pullback of the reference density $\phi$ through $T^{-1}$.
Clearly the pullback only exists when $T$ is monotonic, i.e. has a positive definite Jacobian, and has continuous first derivatives.
Not all maps satisfy these conditions, so we define a smaller space of
maps, $\mathcal{T}^\uparrow \subset \mathcal{T}$, { where $\mathcal{T}$ which contains all
invertible maps. It is an intractable problem to find the
  Knothe–Rosenblatt rearrangement, and so we numerically approximate
  it by formulating an optimisation problem to find the map $\tilde{T}
\in \mathcal{T}^\uparrow$ which most closely maps the target density
to the reference density.}


As in previous work in \cite{parno2018transport}, we can ensure
invertibility if we restrict the map to be {lower triangular\cite{bogachev2005triangular}}, i.e. $\tilde{T} \in \mathcal{T}^{\ltri}\subset\mathcal{T}^\uparrow$. {There are a number of different
  approaches to guaranteeing monotonicity\cite{y.17:_handb_uncer_quant}, but here we follow
  the approach taken in \cite{parno2018transport}.} This lower triangular map has the form,
\begin{equation}
	\tilde{T}(\theta_1, \dots, \theta_n) = \begin{bmatrix} T_1(\theta_1) \\ T_2(\theta_1, \theta_2) \\ \vdots \\
		T_n(\theta_1, \dots, \theta_n) \end{bmatrix},
\end{equation}
where $T_i\colon \mathbb{R}^i \to \mathbb{R}$. 

\subsection{The optimisation problem}
Our aim is now to find the lower triangular map $\tilde{T} \in
\mathcal{T}^{\ltri}$ such that the difference between the target
density and the pullback of the reference density is minimised. As in
\cite{parno2018transport}, we choose the cost function
to be the Kullback-Leibler (KL) divergence between the posterior density and the pullback density,
\begin{equation}
	D_\text{KL}(\pi\|\tilde{\pi}) =
		\mathbb{E}_\pi\left[\log\left(\frac{\pi(\theta)}{\tilde{\pi}(\theta)}\right)\right].
\end{equation}
This divergence results in some nice properties which we will explore in the following derivation. The KL divergence is not a true metric since it is not symmetric, however it is commonly used to measure the distance between probability distributions due to its relatively simple form, and because it provides a bound for the square of the Hellinger distance by Pinsker's inequality~\cite{pinsker1960information},
\begin{equation}
	D_{KL}(p\|q) \geq D_H^2(p,q),
\end{equation}
which is a true distance metric on probability distributions.
Given the form of the pullback in Equation~\eqref{eq:pullback}, now taken through an approximate map $\tilde{T}$, the divergence becomes
\begin{equation}
	D_\text{KL}(\pi\|\tilde{\pi}) = \mathbb{E}_\pi\left[\log\pi(\theta) - \log\pi_r(\tilde{T}(\theta)) -
		\log\left|J_{\tilde{T}}(\theta)\right|\right].
\end{equation}
We note that the posterior density is independent of $\tilde{T}$, and so it is not necessary for us to compute it when optimising this cost function. This expression is a complicated integral with respect to the target distribution, for which the normalisation constant is unknown. However this is exactly the scenario for which we would turn to MCMC methods for a solution.

To find the best coupling, $\tilde{T} \in \mathcal{T}^{\ltri}$, we solve the optimisation problem,
\begin{equation}
	\tilde{T} = \arg\min_{T \in \mathcal{T}^{\smallltri}} \mathbb{E}_\pi\left[-\log\pi_r(T(\theta)) -
		\log\left|J_T(\theta)\right|\right]
\end{equation}
which has a unique solution since the cost function is convex. We also include a regularisation term, which is required for reasons which will become clear later. The optimisation problem now takes the form
\begin{equation}\label{eq:gen_map_optim}
	\tilde{T} = \arg\min_{T\in\mathcal{T}^{\smallltri}} \left[
		 \mathbb{E}_\pi\left[-\log\pi_r(T(\theta)) -
		\log\left|J_T(\theta)\right|\right] + \beta\mathbb{E}(T(\theta)- \theta)^2 \right].
\end{equation}
{The expectation in \eqref{eq:gen_map_optim} can be
approximated by using a Monte Carlo approximation. The form of the regularisation term promotes maps which are
closer to the identity. The parameter $\beta>0$ does not need to be tuned, as experimentation has shown that the choice
$\beta=1$ is sufficient for most problems. Note that this
regularisation term is necessary to avoid overfitting of the transport
map to what is often an incomplete and poorly converged sample from
the posterior in the initial phases. Smaller values of $\beta$ may lead to maps which
hinder further exploration of the state space beyond that already
represented in the sample used to construct the map.}

{Moreover, we can exploit the lower triangular
  structure, for example in the evaluation of 
\begin{equation}\label{eqn:separable_jacobian}
	\log\left|J_{\tilde{T}}(\theta)\right| = \sum\limits_{i=1}^d \! \log \partial_{\theta_i} \tilde{T}_i(\theta),
\end{equation}
where we note that this term is separable in terms of the dimension
$i$. Similarly, inverting $\tilde{T}$ at a point $r$ is simplified by the lower triangular structure of the map.

As in \cite{parno2018transport}, we parameterise each component of the
map $\tilde{T}_i$ with a multivariate polynomial expansion,
\begin{equation}\label{eq:map_defn}
	\tilde{T}_i(\theta; \gamma_i) = \sum\limits_{\mathbf{j}\in\mathcal{J}_i} \!
\gamma_{i,\mathbf{j}}\psi_\mathbf{j}(\theta).
\end{equation}
built from
a family of orthogonal polynomials, $\mathcal{P}(\theta)$, not
necessarily orthogonal with respect to the target distribution. The parameter $\gamma_i \in \mathbb{R}^{M_i}$ is a vector of coefficients. Each component of $\gamma_i$ corresponds to a basis function
$\psi_\mathbf{j}$, indexed by the multi-index $\mathbf{j} \in
\mathcal{J}_i \subset \mathbb{N}_0^i$. 
Here we stick with polynomials of total order $p$ since we work with
low dimensional problems with the ETAIS algorithm. More details about
the structure of the maps can be found in  \cite{parno2018transport}.}

\subsection{Implementation of the optimisation problem}\label{sec:transport_implementation}

{We now discuss how we can evaluate the cost function in
Equation~\eqref{eq:gen_map_optim}. In \cite{parno2018transport}, this
expectation is approximated using an Monte Carlo estimator with samples from
a Metropolis-Hastings method, which have equal weighting.
Here we aim to build a map from
samples from an importance sampling scheme.} Such samples no longer
carry equal weight, and as such the Monte Carlo estimator becomes
\begin{equation}\label{eqn:TM_full_cost}
	C(T) = \frac{1}{\bar{w}}\sum\limits_{i=1}^d \! \sum\limits_{k=1}^K
		w_k \left[-\log\pi_r(T_i(\theta^{(k)})) -
			\log\left|\frac{\partial T_i}{\partial \theta_i}(\theta^{(k)})\right| + \beta(T_i(\theta^{(k)})-\theta^{(k)})^2\right],
\end{equation}
where $w_k$ are the weights associated with each sample $\theta^{(k)}$, and $\bar{w}$ is the sum of
all these weights. Optimisation of this cost function results in a map from $\pi$ to some reference density $\pi_r$. By choosing the reference density to be a Gaussian density, we can simplify this expression greatly. Substitution of the Gaussian density into Equation~\eqref{eqn:TM_full_cost} leads to
\begin{equation}\label{eqn:TETAIS_objective}
	C(T) = \frac{1}{\bar{w}}\sum\limits_{i=1}^d \! \sum\limits_{k=1}^K
		w_k\left[\frac{1}{2}T_i^2(\theta^{(k)}) - \log\frac{\partial
		T_i}{\partial\theta_i}(\theta^{(k)}) + \beta(T_i(\theta^{(k)})-\theta^{(k)})^2\right],
\end{equation}

Note that since we assume that the map is monotonic, the derivatives of each component are
positive and so this functional is always finite. In practice it is infeasible to enforce this condition across the entire parameter space. We instead enforce this condition by ensuring that the derivatives are positive at each sample point. This means that when we sample away from these support points while in reference space, it is possible to enter a region of space where the map is not monotonic.

We now return to the structure of the map components given in equation
\ref{eq:map_defn}. Since the basis functions are
fixed, the optimisation problem in \eqref{eq:gen_map_optim} is really over the map components $\bar{\gamma} = (\gamma_1, \dots,
\gamma_d)$ where $\gamma_i \in \mathbb{R}^{M_i}$. Note that $C(T)$ is the sum of $d$ expectations, and these expectations each only concern one dimension. Therefore we can rewrite \eqref{eq:gen_map_optim} as $d$ separable optimisation problems.
\begin{align}\label{eq:gamma_map_optim}
	&\arg\min_{\gamma_i\in\mathbb{R}^{M_i}} \frac{1}{\bar{w}}\sum\limits_{k=1}^K
		w_k \left[\frac{1}{2}T_i^2(\theta^{(k)}; \gamma_i) - \log\frac{\partial T_i}{\partial\theta_i}(\theta^{(k)}; \gamma_i) + \beta(T_i(\theta^{(k)};
		\gamma_i)-\theta^{(k)})^2\right], \\
	&\text{subject to} \quad \frac{\partial T_i}{\partial\theta_i}(\theta^{(k)};
		\gamma_i) > 0 \ \text{for all}\ k=1,\dots,K,\ i=1,\dots,d.
		\notag
\end{align}
The sum in Equation~\eqref{eq:map_defn} is an inner
product between the vector of map coefficients, and the evaluations of the basis function at a
particular $\theta^{(k)}$. If we organise our basis evaluations into two matrices,
\begin{equation}
	(F_i)_{k,\mathbf{j}} = \psi_\mathbf{j}(\theta^{(k)}), \quad \text{and} \quad (G_i)_{k,\mathbf{j}} =
\frac{\partial\psi_\mathbf{j}}{\partial\theta_i}(\theta^{(k)}),
\end{equation}
for all $\mathbf{j}
\in \mathcal{J}_i^\text{TO}$, and $k = 1,\dots,K$, then we have that
\begin{equation}
	T_i(\theta^{(k)}) = (F_i)_{k\cdot}\gamma_i \quad \text{and} \quad \frac{\partial T_i}{\partial \theta_i}(\theta^{(k)}; \gamma_i) = (G_i)_{k\cdot}\gamma_i,
\end{equation}
so \eqref{eq:gamma_map_optim} becomes
\begin{align}\label{eq:blas_map_optim}
	&\arg\min_{\gamma_i\in\mathbb{R}^{M_i}}
          \frac{1}{2}(F_i\gamma_i)^\top W (F_i\gamma_i) -
		{\bf w}^\top\log(G_i\gamma_i) + \frac{\beta}{\bar{w}}\sum\limits_{k=1}^K \!
		w_k(F_i\gamma_i-\theta^{(k)})^\top(F_i\gamma_i-\theta^{(k)}), \\
	&\text{subject to} \quad G_i\gamma_i > 0. \notag
\end{align}
In this expression, the vector ${\bf w} = [w_1, w_2, \ldots,
w_K]^\top$ is the vector of the weights, $W$ is the diagonal matrix $W
= \text{diag}(w)$ and $\log(G_i\gamma_i)$ is to be
evaluated element-wise. As more importance samples are made, new rows can be appended to the
$F_i$ and $G_i$ matrices, and $F_i^\top W F_i$ can be efficiently updated via the addition of rank-1 matrices.

The regularisation term in Equation~\eqref{eq:blas_map_optim} can be approximated using Parseval's identity,
\begin{equation}
	\frac{1}{\bar{w}}\sum\limits_{k=1}^K \! w_k
        (F_i\gamma_i-\theta^{(k)})^\top(F_i\gamma_i-\theta^{(k)})
        \xrightarrow[K \to \infty]{}
		\int_{\mathbb{R}^n} |T(\theta)-\theta|^2 \text{d}\pi_\theta =
		\sum\limits_{\mathbf{j}\in\mathcal{J}_i^\text{TO}} (\gamma_{i,\mathbf{j}}-\iota_\mathbf{j})^2,
\end{equation}
where $\iota$ is the vector of coefficients for the identity map. This is of course only true when
the polynomial family $\mathcal{P}(\theta)$ is chosen to be orthonormal with respect to $\pi_\theta$; however this
approximation prevents the map from collapsing onto a Dirac when the expectation is badly approximated by a small number of samples.

These simplifications result in the efficiently implementable, regularised optimisation problem for
computing the map coefficients, 
\begin{align}\label{eq:weighted_map_optim}
	&\arg\min_{\gamma_i\in\mathbb{R}^{M_i}} \frac{1}{2\bar{w}}\gamma_i^\top F_i^\top WF_i\gamma_i -
		\frac{w^\top}{\bar{w}}\log(G_i\gamma_i) + \beta\|\gamma_i-\iota\|^2, \\
	&\text{subject to} \quad G_i\gamma_i > 0. \notag
\end{align}
This optimisation problem can be efficiently solved using Newton iterations. It is suggested
in~\cite{parno2018transport} that this method usually converges in around 10-15 iterations, and we
have seen no evidence that this is not a reasonable estimate. When calculating the map several times
during a Monte Carlo run, using previous guesses of the optimal map to
seed the Newton algorithm 
results in much faster convergence, usually taking only a couple of iterations to satisfy the stopping
criteria.

The Hessian takes the
form
\begin{equation}\label{eqn:TETAIS_hessian}
	HC_i(\gamma_i) = \frac{1}{\bar{w}}\left[F_i^\top WF_i + G_i^\top
		W\text{diag}([G_i\gamma_i]^{-2})G_i\right] + \beta I,
\end{equation}
where $[G_i\gamma_i]^{-2}$ is to be taken element-wise, and $I$ is the $M_i\times M_i$
identity matrix. The first derivative of $C_i(T)$ is
\begin{equation}
	\nabla C_i(\gamma_i) = \frac{1}{\bar{w}}\left[F_i^\top WF_i\gamma_i - G_i^\top
		W[G_i\gamma_i]^{-1}\right] + \beta(\gamma_i - \iota),
\end{equation}
again $[G_i\gamma_i]^{-1}$ is taken element-wise.

\section[Transport map MCMC]{Transport map usage in ETAIS
  and other importance sampling algorithms}\label{sec:TETAIS}

Given importance samples from the target distribution, we have demonstrated how to construct an approximate transport map from the
target measure to a reference measure. We now consider how to
implement an ensemble importance sampling algorithm which uses
these maps to propose new states. In \cite{parno2018transport} it was
shown how approximate transport maps can be used to accelerate
Metropolis-Hastings methods, with the map being periodically updated
through the samples produced from the target measure. Convergence of this
adaptation is shown in~\cite{parno2018transport}. In this Section, we
will show how similarly, these maps can be used to construct highly
efficient importance sampling schemes.

In particular, we will show how we can use the transport map derived in Equation~\eqref{eq:weighted_map_optim} to
design a proposal scheme for the ETAIS algorithm. In this case we have a choice in how to proceed; we
propose new samples on reference space and resample on target space, or we both propose and resample on reference space, mapping onto target space to output the samples. The first option allows us to reuse much of the framework
from the standard ETAIS algorithm and in the numerics later we see that this performs better than
both the Transport MH algorithm, and the standard ETAIS algorithm. The second option requires some
restructuring, but can result in improved performance from the resampler.




\begin{table}
\begin{algorithm}[H]
\DontPrintSemicolon
\BlankLine
Initialise state $\theta^{(1)}_i = \theta_0$, \quad $i = 1,\dots,M$.\;
Initialise map $\bar{\gamma}^{(1)} = \iota$.\;
\For{$k \leftarrow 1, \dots, N-1$}{
	Compute $r_i = \tilde{T}(\theta^{(k)}_i; \bar{\gamma}^{(k)})$, \quad $i = 1,\dots,M$.\;
	Sample {$\hat{r}_i \sim q_r(\cdot; r_i, \beta)$.\;
	Invert $\hat{\theta}_i^{(k)} = \tilde{T}^{-1}(\hat{r}_i; \bar{\gamma}^{(k)})$.\;
	Calculate:
	\begin{equation}
		w_i^{(k)} =
                \frac{\pi(\hat{\theta}_i^{(k)})}{\left(\sum_{j=1}^M \!
                    {q_r(\hat{r}_i; r_j, \beta)}\right)|J_{\tilde{T}}(\hat{\theta}_i^{(k)};\bar{\gamma}^{(k)})|}.
	\end{equation}

	Resample $r^* \leftarrow \|w^{(k)}\|^{-1}\sum\limits_{j=1}^M \! w_j^{(k)}\delta_{\hat{r}_j}(\cdot)$.\label{algline:TETAIS_resample}\;}
	Invert $\theta^{(k+1)}_i = \tilde{T}^{-1}(r^*_i)$.\;
	\eIf{$k\ \text{mod}\ K_U = 0$ and $k < K_\text{stop}$}{
		\For{$i \leftarrow 1, \dots, n$}{
			Solve \eqref{eq:weighted_map_optim} with $\{(w^{(1)},\hat{\theta}^{(1)}), \dots, (w^{(k+1)},\hat{\theta}^{(k+1)})\}$
				and update $\gamma_i^{(k+1)}$.\;
		}
	}{
		$\bar{\gamma}^{(k+1)} = \bar{\gamma}^{(k)}$.\;
	}
}
\caption{ETAIS algorithm with adaptive transport map. Option 2.\label{alg:TransportETAIS2}}
\end{algorithm}
\end{table}

The second option is given in Algorithm~\ref{alg:TransportETAIS2}. We
denote the ensembles of states in target space $\theta^{(k)} =
\{\theta^{(k)}_1,\dots,\theta^{(k)}_M\}$, and the states in the
reference space, {$r = \{r_1,\dots,r_M\}$}, where $M$ is the ensemble
size. Similarly, the proposal states are denoted $\hat{r} =
\{\hat{r}_1,\dots,\hat{r}_M\}$ and $(w^{(k)}, \hat{\theta}^{(k)}) =
\{(w^{(k)}_1, \hat{\theta}^{(k)}_1),\dots,(w^{(k)}_M,
\hat{\theta}^{(k)}_M)\}$, where these {pairs constitute the
  weighted samples} from the target distribution. As in the standard version of the ETAIS algorithm, we use the deterministic mixture weights.

The second option, is similar to
the first except on Line~\ref{algline:TETAIS_resample}, where rather
than resampling in target space we resample in reference
space. Assuming a good transport map, the pushforward of the target
measure onto 
reference space has no strong correlations, and the
Gaussian marginals are easy to approximate with fewer ensemble
members. This means that the resampling step will be more efficient in
moderately higher dimensions, which we discuss in
Section~\ref{sec:TETAIS_higher_dim}.

{The choice of ensemble size can be problem dependent, but it is quite clear when the ensemble size is
  too small, because the proposal distribution poorly represents the
  targets, and the variance of the weights is high. The ensemble size
  can also be made to be adaptive, so that when the variances grow too
high the ensemble size can increase. The ensemble size needs to be
controlled since the cost of the resampling step is often polynomial
in the ensemble size.}

\section[Convergence of transport MCMC]{Convergence of the transport proposal based MCMC algorithms}\label{sec:conv}

\begin{figure}
\centering
\subfigure[Marginal density function for $\theta_1$.]{\includegraphics[width=0.31\textwidth]{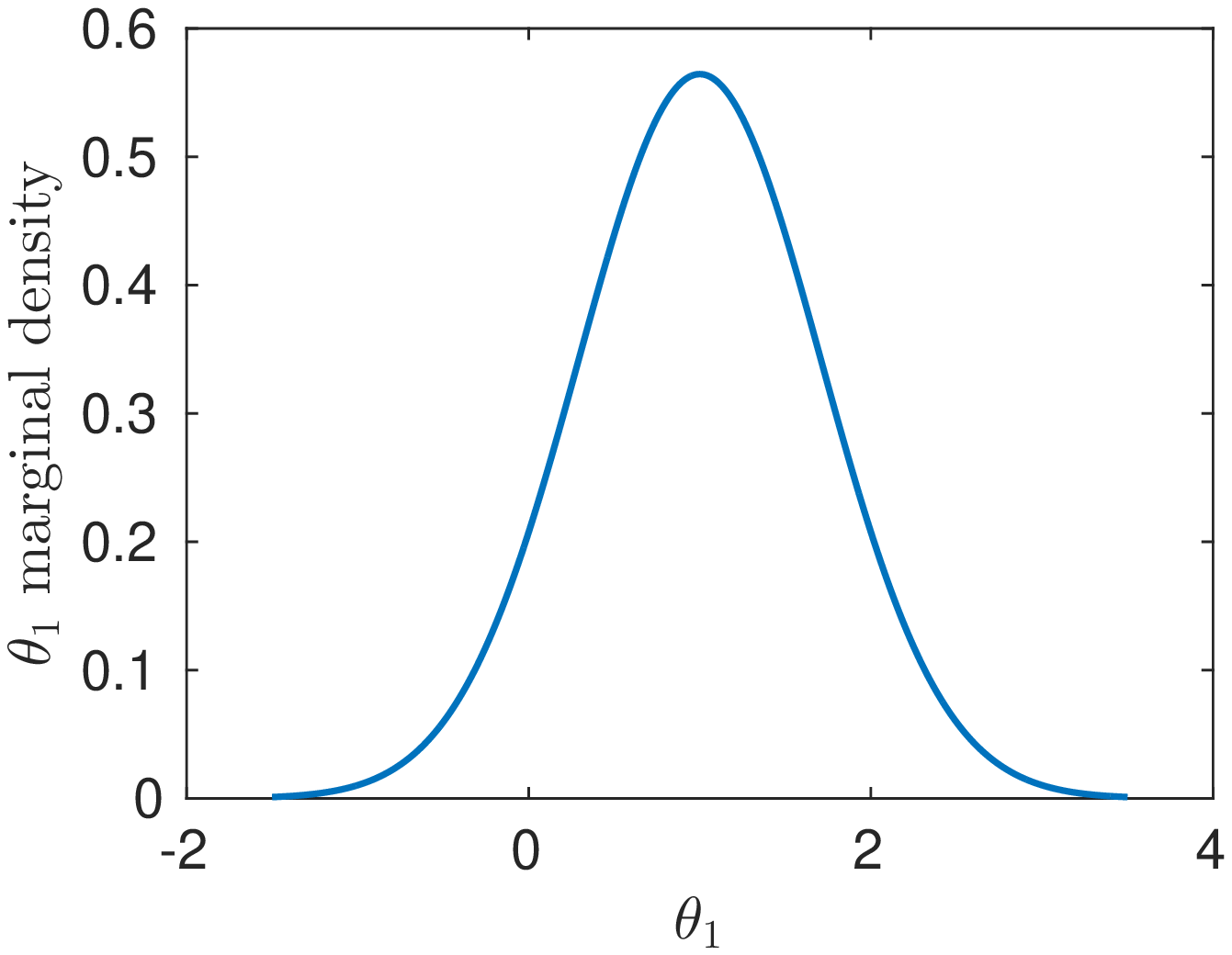}}
\subfigure[Marginal density function for $\theta_2$.]{\includegraphics[width=0.31\textwidth]{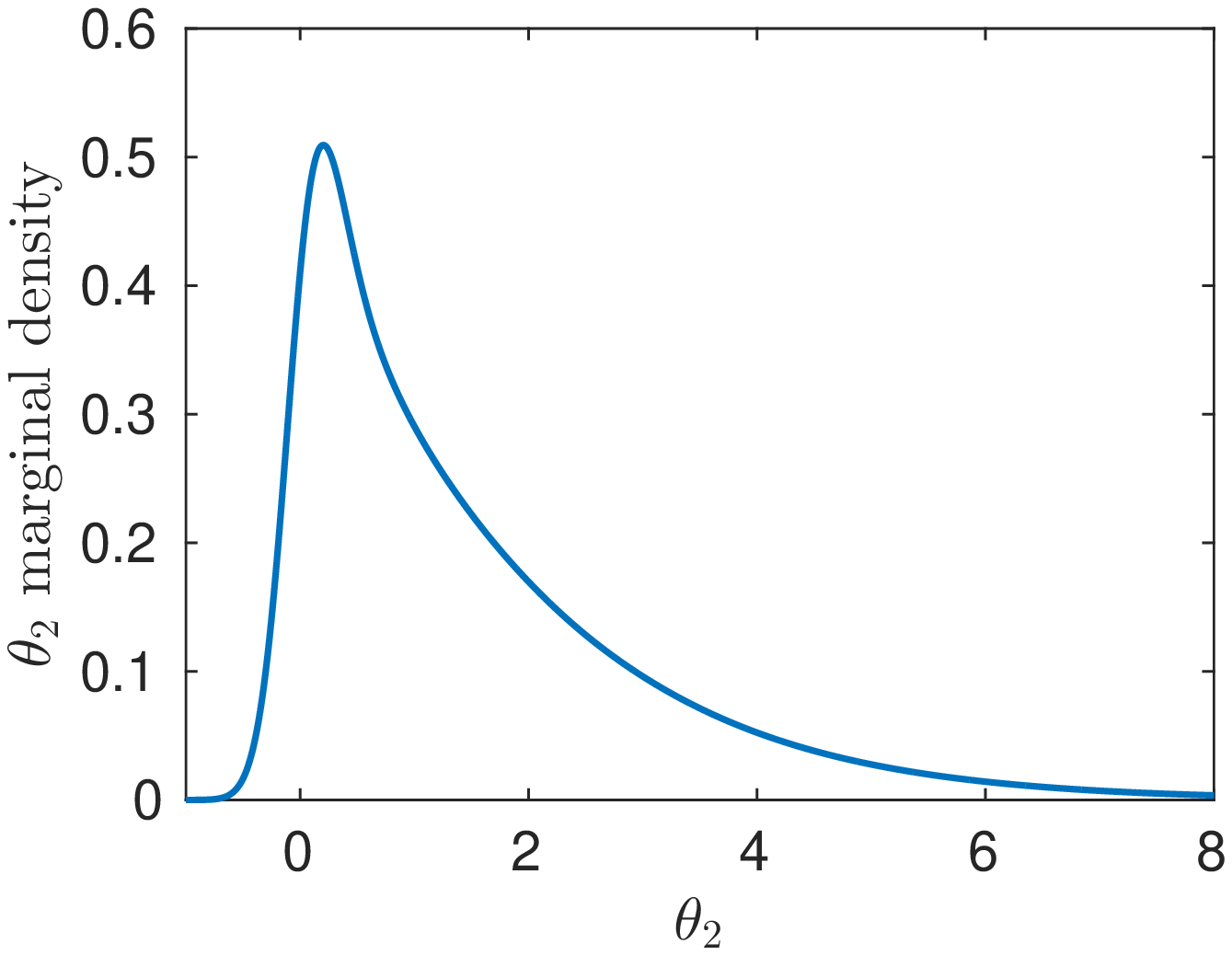}}
\subfigure[Contour plot for Rosenbrock density.]{\includegraphics[width=0.31\textwidth]{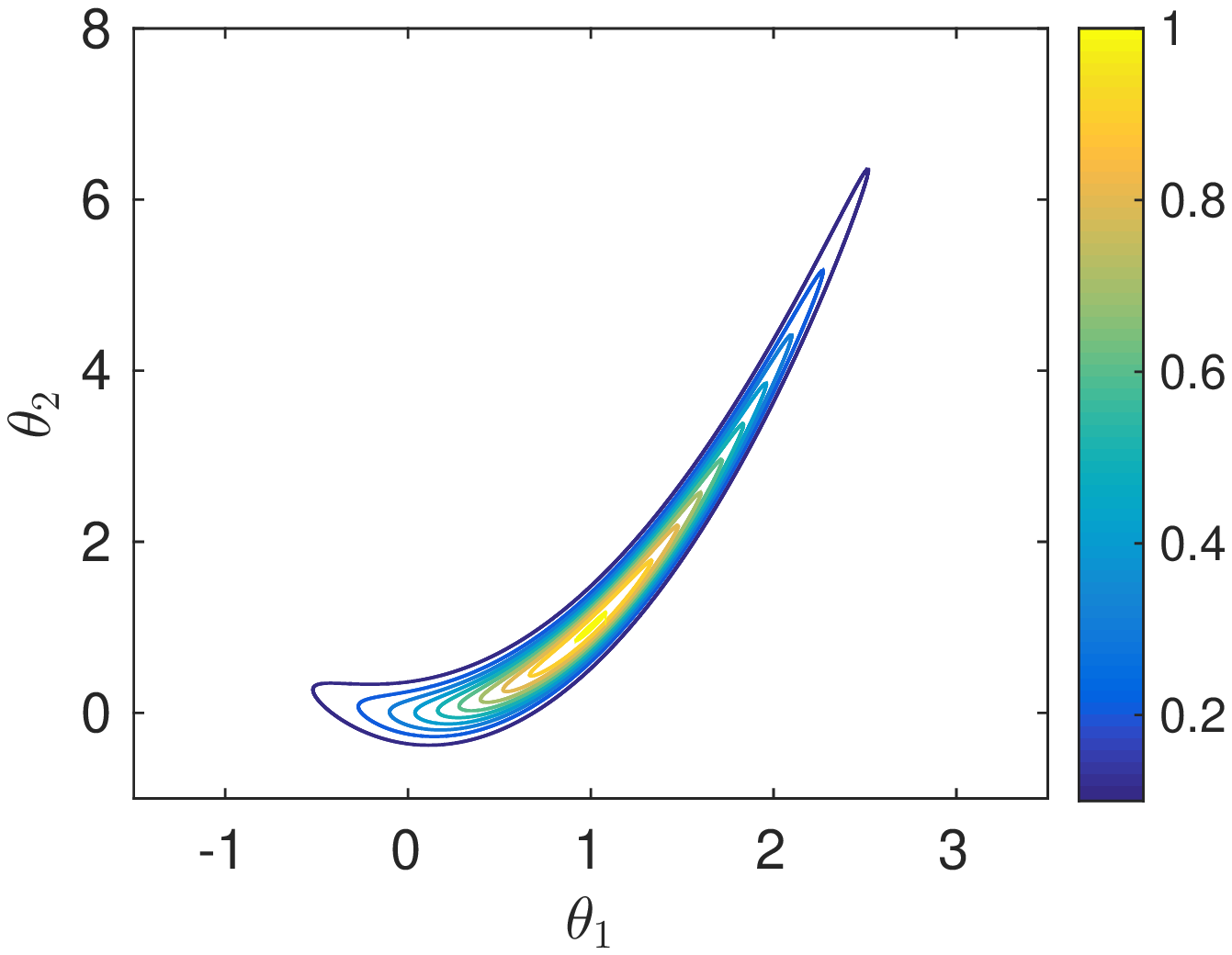}}
\caption{Visualisation of the Rosenbrock density as given in Equation~\eqref{eqn:R2_repeat}.}
\label{fig:R2_posterior}
\end{figure}

In this Section we study the convergence of the transport based
proposal distributions which we have described in
Section~\ref{sec:TETAIS}.
We take as a test problem the ubiquitous Rosenbrock banana-shaped
density. This target density is given by
\begin{equation}\label{eqn:R2_repeat}
	\pi(\theta) = \frac{\sqrt{10}}{\pi}\exp\left\{ -(1 - \theta_1)^2 - 10(\theta_2 - \theta_1^2)^2 \right\}.
\end{equation}
A contour plot of the target density is given in Figure~\ref{fig:R2_posterior}.
This problem is challenging to sample from since it has a highly peaked and curved ridge, and is often used
as a test problem in optimisation and MCMC communities.

\subsection{Implementation details}

Before looking at the performance of the algorithms, we demonstrate some properties of the transport maps we will be using. We draw 1 million samples from the density \eqref{eqn:R2_repeat}, and use this sample in the framework of Section~\ref{sec:map} to build a transport map. We use this map to pushforward the original sample onto the reference space, where we will be able to see how well the map has performed at converting the original sample to a standard Gaussian. We then pull the sample back on to target space using the inverse map to check that our map is invertible and well behaved.

For this example, we use an index set of total order 3 with monomial
basis functions. It is important that total order is an odd number,
since otherwise the map will not be surjective. This results in a map of the form
\begin{equation}
	T(\theta_1, \theta_2) = \begin{bmatrix} T_1(\theta_1) \\ T_2(\theta_1, \theta_2) \end{bmatrix},
\end{equation}
where
\begin{align}
		T_1(\theta_1) &= \gamma_{1,1} + \gamma_{1,2}\theta_1 + \gamma_{1,3}\theta_1^2 + \gamma_{1,4}\theta_1^3, \\
		T_2(\theta_1, \theta_2) &= \gamma_{2,1} + \gamma_{2,2}\theta_1 + \gamma_{2,3}\theta_1^2 + \gamma_{2,4}\theta_1^3
					+ \gamma_{2,5}\theta_2 + \gamma_{2,6}\theta_1\theta_2 \\
				 & \qquad \quad + \gamma_{2,7}\theta_1^2\theta_2 + \gamma_{2,8}\theta_2^2 + \gamma_{2,9}\theta_1\theta_2^2 +
					 \gamma_{2,10}\theta_2^3.
\end{align}
Clearly even with only basis functions of total order 3, we have a large number of unknowns in our optimisation problem, $\bar{\gamma} \in \mathbb{R}^{14}$. If we were to increase the dimension of $\theta$ further, we would need to reduce the number of terms we include in the expansion by, for example, removing all the ``cross'' terms. This reduces the quality of our map, but since we only require an approximate map we can afford to reduce the accuracy.

\begin{figure}[htpb]
\centering
\subfigure[Original sample $\theta$ from MH-RW algorithm.]{\includegraphics[width=0.31\textwidth]{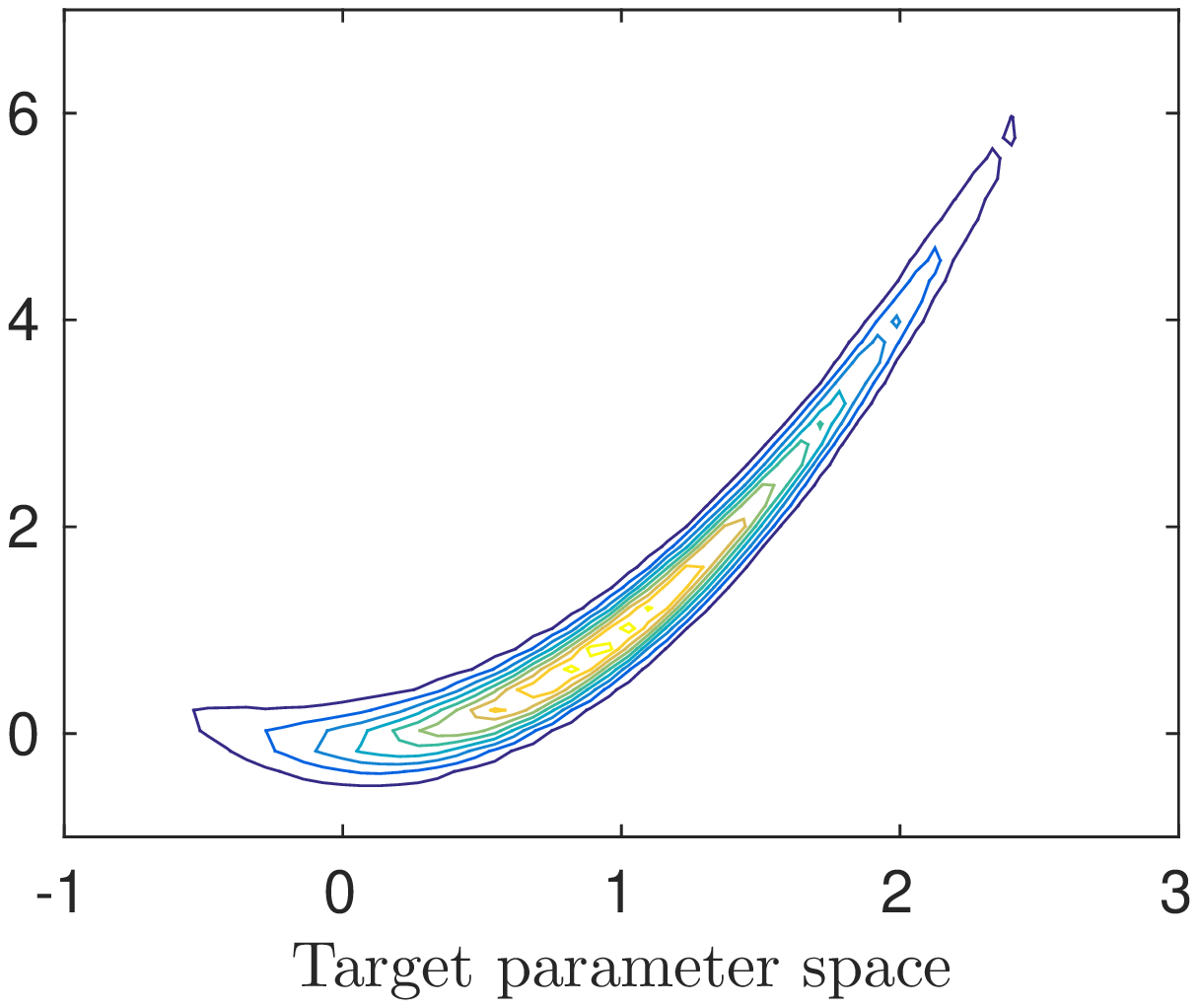}}\quad
\subfigure[Pushforward of $\theta$ onto reference space.]{\includegraphics[width=0.31\textwidth]{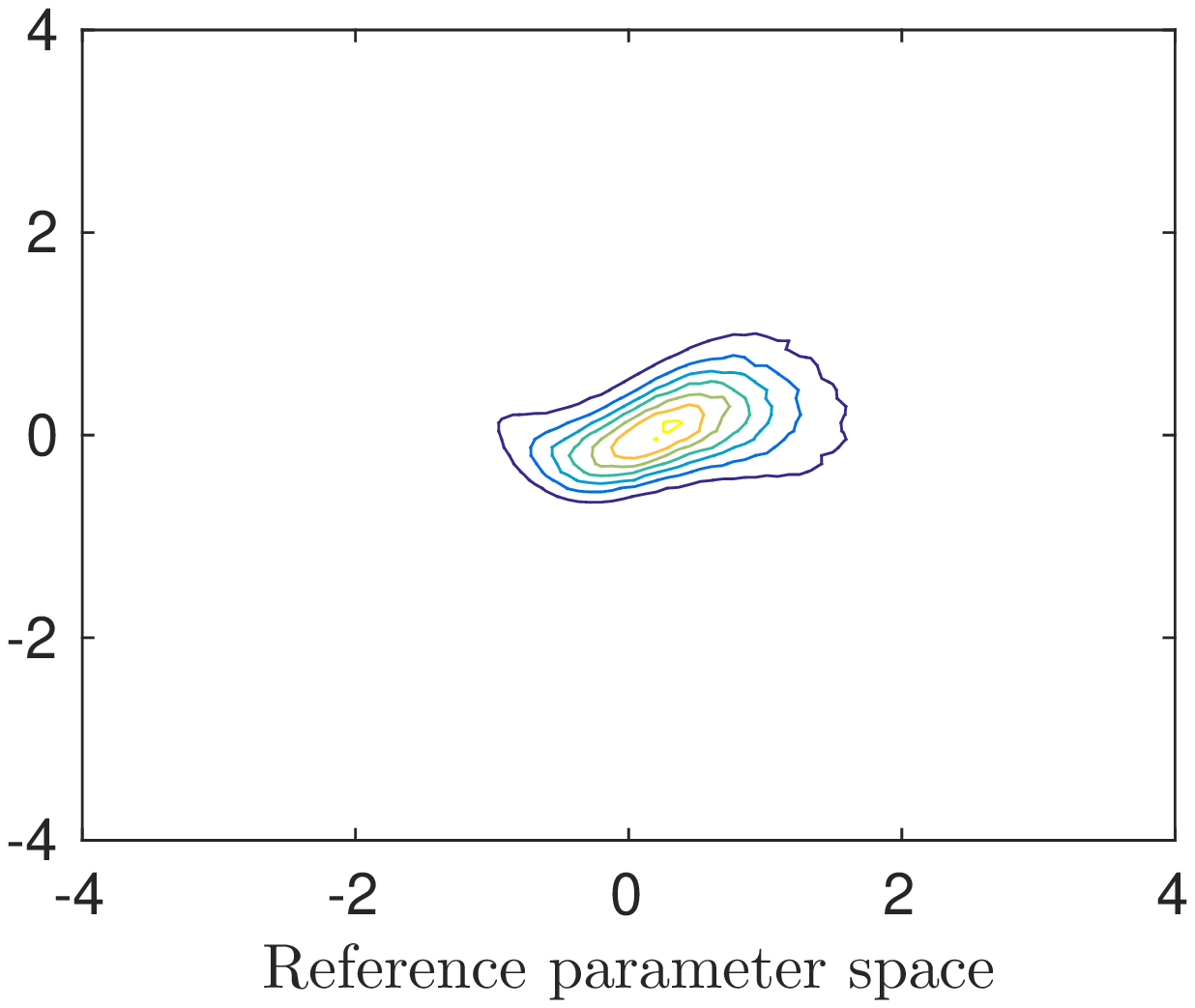}}\quad
\subfigure[Pullback of reference sample onto target space.]{\includegraphics[width=0.31\textwidth]{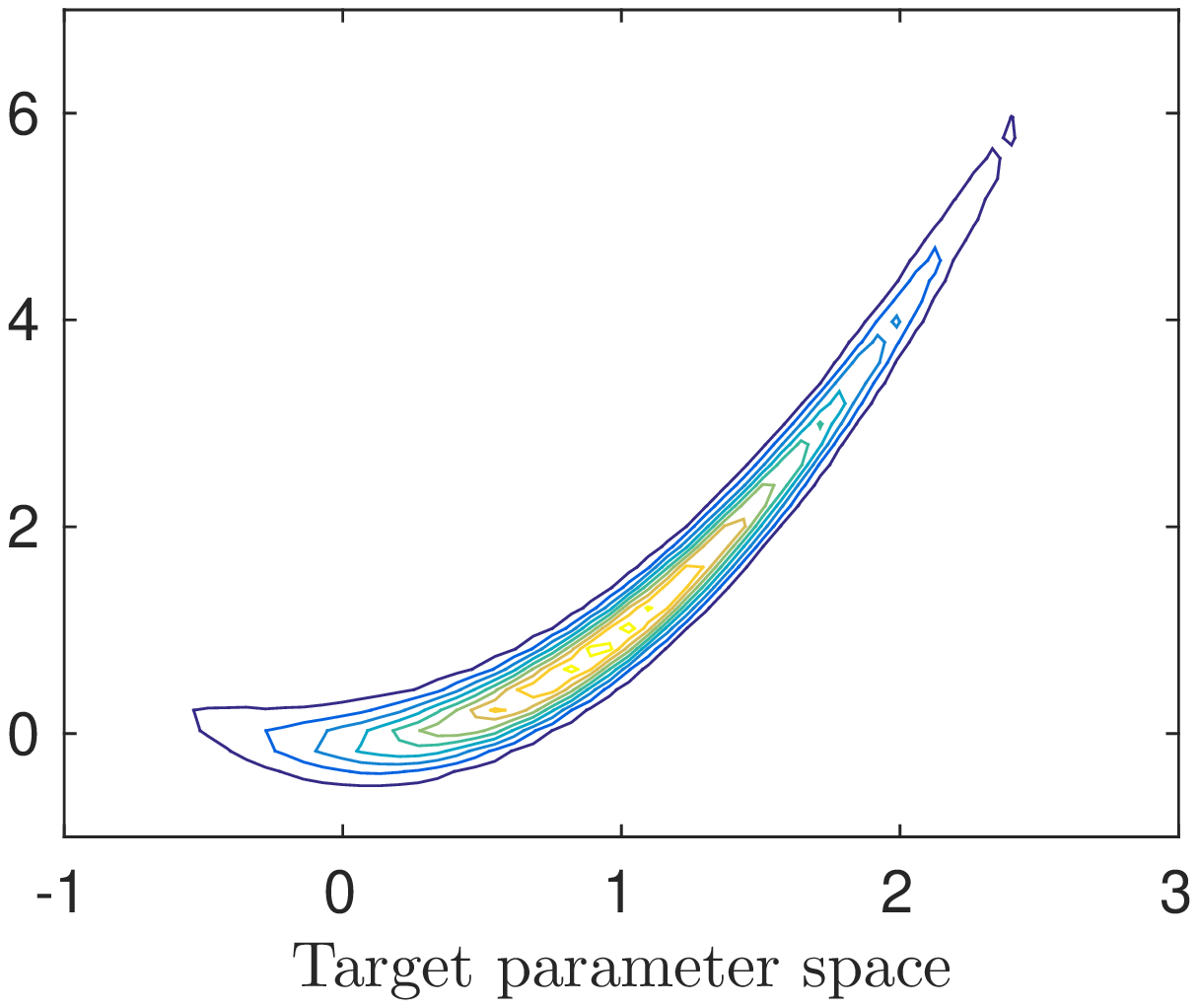}}
\caption{The effect of the approximate transport map $\tilde{T}$ on a
  sample from the Rosenbrock target density as described in Equation~\eqref{eqn:R2_repeat}.}
\label{fig:R2_transport}
\end{figure}

Figure~\ref{fig:R2_transport} shows the output of the approximate
transport map. Even though we have truncated the infinite expansion in
the monomial basis down to 4 and 10 terms in respective dimensions,
the pushforward of the sample is still a unimodal distribution
centred at the origin with standard deviation 1. As one moves out into
the tails of the reference density more non-Gaussian features are
clearly visible. However, overall, the pushforward of the target
density does not look a challenging one to sample from, with even
relatively simple MCMC methods such as Metropolis-Hasting random walk (MH-RW). The pullback from
reference space, in Figure~\ref{fig:R2_transport}, is an exact match
of the original sample since we have not perturbed the sample in
reference space. This inversion is well defined in the sampling
region, although not necessarily outside it\cite{parno2018transport}.

{For more challenging posterior distributions, the pushforward
  through the numerical
approximation of the optimal map often leads to a distribution on
reference space which still has complex structure. It has previously
been proposed that in more favourable scenarios, the optimal approach
is to use independent Gaussian samples on reference space, pulled back
through the inverse map. However, when the pushforward of the
posterior differs greatly from this distribution, as we will see in
later examples, this is not an
efficient approach. Using ETAIS in conjunction with the transport map
allows us to sample efficiently from the pushforward measure on
reference space.}

\subsection[Numerical results]{Numerical results for convergence of
  transport map based algorithms on the Rosenbrock density}
{The performance of Metropolise-Hastings methods and adaptive
importance sampling methods depends on the scaling parameter
$\beta^2$ for the proposal distributions $q(\cdot ;
\cdot, \beta)$. For the RW method, for example, the variance of the
proposal is equal to $\beta^2$.}
We first find the optimal scaling parameters for the individual
algorithms. This is done, as in \cite{cotter2015parallel}, by optimising for the effective sample size in
the ETAIS algorithm, and by tuning the relative $L^2$ error in the MH
algorithm. 
{The relative $L^2$ error of the PDF of the measure can be
  approximated by $E$ where
\begin{equation}\label{eqn:L2_error}
	E^2 = \sum\limits_{i=1}^{n_b}\left[\displaystyle\int_{R_i} \! \pi(s|D) \, \mbox{d}s - vB_i\right]^2 \Big/ \sum\limits_{i=1}^{n_b}\left[\displaystyle\int_{R_i} \! \pi(s|D) \, \mbox{d}s\right]^2,
\end{equation}
where the regions $\{R_i\}_{i=1}^{n_b}$ are the $d$-dimensional
histogram bins, so that $\bigcup_i R_i \subseteq X$ and
$R_i\cap R_j=\emptyset$, $n_b$ is the number of bins, $v$ is the
volume of each bin, and $B_i$ is the value of the $i$th bin. This
metric converges to the standard definition of the relative $L^2$
error as $v\rightarrow 0$.}

There is currently no
guidance on the best way of tuning the MH algorithm with transport map
proposals, although one might expect results similar to the standard MH
results, especially if adaptation of the map is stopped after a given
point. As in the ETAIS algorithm, optimising for the effective sample size might be the best option.

\begin{table}[!ht]
\centering
\begin{tabular}{lrrr}
\toprule
	Statistic \quad / \quad Algorithm & Transport M-H &
                                                           Transport
                                                            ETAIS
                                                            option 1 & Alg. \ref{alg:TransportETAIS2}  \\ \cmidrule(lr){1-4}
	{$\beta_{L^2}$}				 & 1.0e-0 & 1.1e-1 & 3.5e-1 \\
	{$\beta_{\text{ESS}}$}				 & - & 1.0e-1 & 5.2e-1 \\ \cmidrule(lr){1-4}
	Acc. rate							 & 0.23 & - & - \\
	ESS ratio							 & - & 0.62 & 0.71 \\
\bottomrule
\end{tabular}
\caption{{Optimal scaling parameters $\beta$ for the proposal
    kernels with the transport map based
  algorithms applied to $R_1$, optimising for the approximate
    $L^2$ error~\eqref{eqn:L2_error} and
  the effective sample size (ESS) for ETAIS algorithms, and for average
  acceptance rate for MH algorithms.}}
\label{tab:R2_opt_scaling}
\end{table}

The optimal scaling parameters are given in
Table~\ref{tab:R2_opt_scaling}. Here we see that the effective sample
size is much lower than we see in the one-dimensional examples with
the ETAIS algorithms. However, in the Rosenbrock density
\eqref{eqn:R2_repeat} we are dealing with a much more complicated
curved structure, as well as a very slowly decaying tail in
$\theta_2$. From our experiments, we have observed that the standard ETAIS-RW required an ensemble size of $M=500$ to overcome the problems in this density, however the transport map transforms the tails to be more like those of a Gaussian which can be approximated well by a smaller ensemble size of $M=150$.

\begin{figure}[!ht]
\centering
\subfigure[Comparison of the two Transport ETAIS 
options.]{\includegraphics[width=0.45\textwidth]{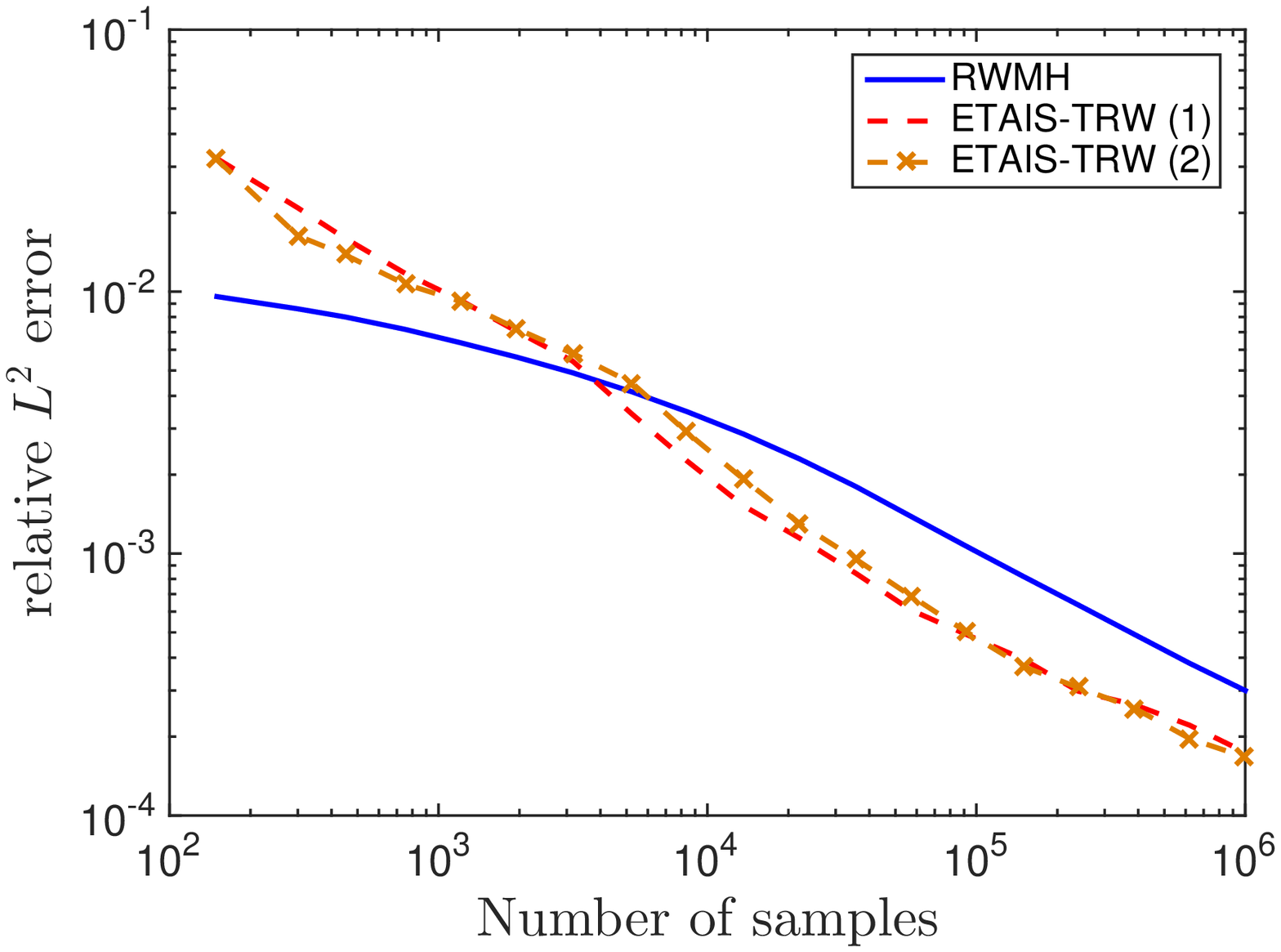}}
\subfigure[Comparison of Transport ETAIS option 2 with the standard algorithms.]{\includegraphics[width=0.45\textwidth]{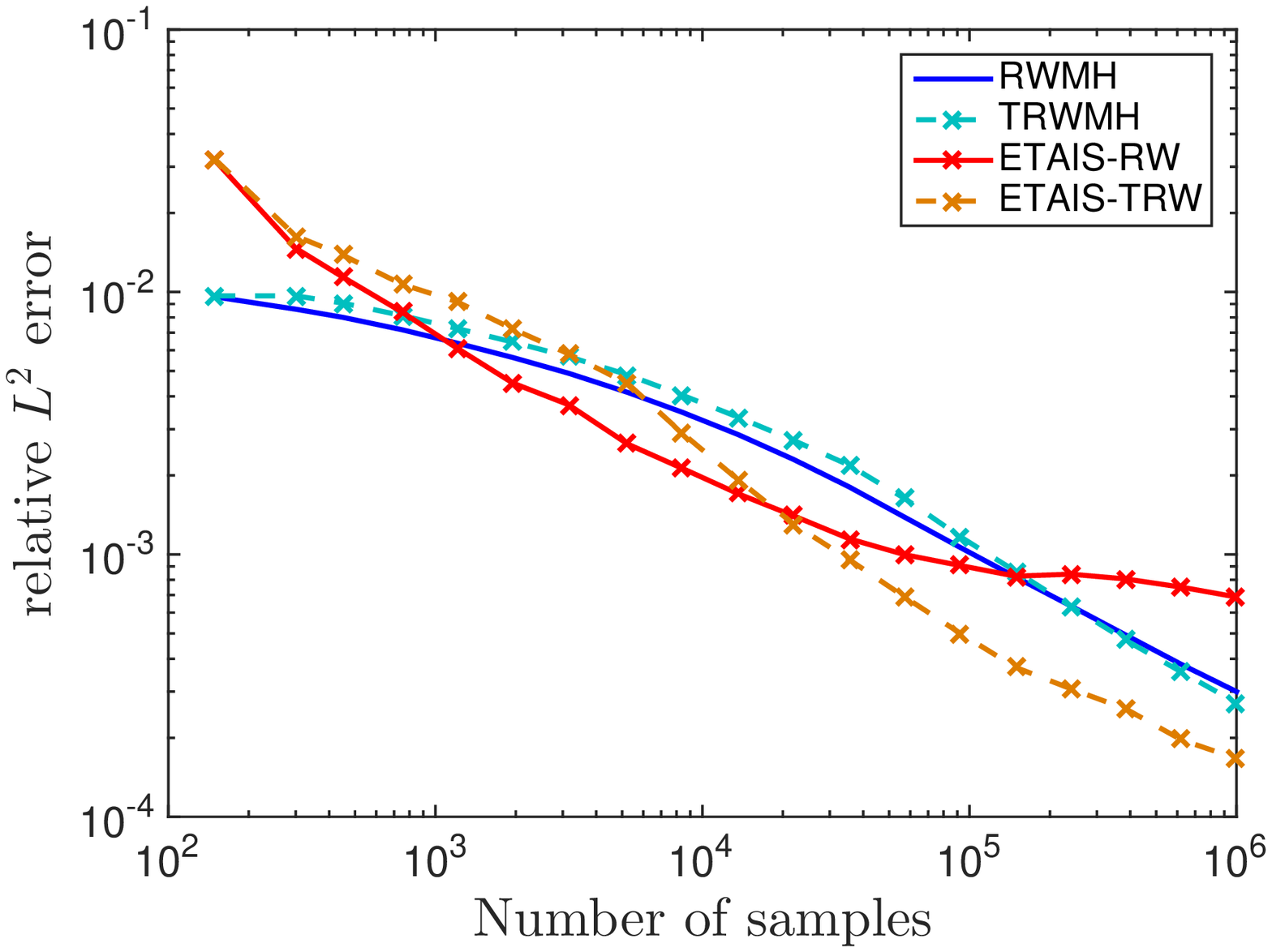}}
\caption{Convergence of algorithms; Transport M-H,
  Transport ETAIS option 1, algorithm \ref{alg:TransportETAIS2} for
  density \eqref{eqn:R2_repeat}. Ensemble size $M=150$, resampling
  performed using the MT algorithm. {Plots show average
    convergence over 32 repeats.}}
\label{fig:R2_l2_convergence}
\end{figure}

The convergence of the three algorithms is displayed in
Figure~\ref{fig:R2_l2_convergence}. {These convergence plots show
  the error as a function of the number of samples output by each
  algorithm. In the case of standard Metropolis-Hastings methods, this
  corresponds to the number of iterations. In the case of ETAIS, this
  corresponds to the number of iterations multiplied by the ensemble
  size.} Figure (a) shows that the two
variations of the transport based ETAIS algorithms converge with
similar rates. The second version, which performs the resampling stage
in reference space rather than target space, has a slightly higher
ESS, and is more stable than option (1). This version also has a
property that we can exploit in Section~\ref{sec:TETAIS_higher_dim}.

Figure~\ref{fig:R2_l2_convergence} (b) compares the performance of
RWMH, transport map-accelerated RWMH, ETAIS-RW and transport
map-accelerated ETAIS-RW. The difference in performance between RWMH
and its accelerated version is relatively negligible, in part because
this version of the Rosenbrock density is not overtly tortuous. The story
is different for the ETAIS algorithms. The slowly decaying tails of
the Rosenbrock cause a problem for ETAIS since these are not
well-represented in the proposal distributions, leading to high weight
variance and poor convergence. A much larger
ensemble would be required to ensure a stable ETAIS implementation for this
problem, which leads to high overhead costs for the resampling step
and evaluation of the importance weights. The transport map-accelerated
ETAIS, on the other hand, is stabilised by the map, which transforms the proposal
distributions in order to adequately approximate the slowly decaying
tails of the target. This stabilisation allows us to benefit from the
improved convergence of the ETAIS methodology, with a smaller ensemble
size.


\section{Inverse Problems for Multiscale Stochastic Chemical Reaction
  Networks}\label{sec:multi}
{In this Section, we briefly introduce the idea of a continuous
  time Markov chain model for stochastic reaction networks. We consider the formulation of the Bayesian inference of
  the reaction parameters given exact data. In particular we consider
  the case where only slow variables in the system are observable, and
  hence the true likelihood function is intractable. In this case, the
  likelihood can be approximated using multiscale methods. The data in
  these scenarios is often insufficient to identify all of the
  parameters, and this can lead to posterior distributions which are
  heavily concentrated close to lower dimensional manifolds, which are
  challenging to sample from, and which motivated the development of
  the methodology outlined in this paper.}

We consider chemical reaction networks of $N_s$ chemical species $\{S_j\}_{j=1}^{N_s}$,
with population numbers given by $X(t) = [X_1(t), X_2(t), \ldots, X_{N_s}(t)]^\top \in
\mathbb{N}_0^{N_s}$ reacting in a small reactor, through $N_r$ different reaction
channels. When population numbers of one or more of the chemical
species in the reactor are small, as is the case with chemical
reactions occurring within living cells, the sporadic and discrete way
in which reactions occur can not be well modelled by deterministic
continuous models, such as ODEs. In such a situation, we may wish to
model the dynamics through a discrete stochastic model, such as a
continuous time Markov chain.

For each reaction $R_j$, for $j = 1,2,\ldots N_r$, there is a
propensity or hazard function $\alpha_j(X(t))$ which indicates how
likely that reaction is to fire, defined by
\begin{equation}\alpha_j(X(t)) = \lim_{dt \to 0} \mathbb{P}(\text{Reaction $R_j$ in
    the time interval  } \, s \in [t, t+ dt] ).\end{equation} 

Following each reaction $R_j$ there is an instantaneous change in the
current state, as the reactants of the reaction are consumed, and the
products produced. This is modelled by a change in the state vector
$X(t) = X(t) + \nu_j$ where $\nu_j$ is the stoichiometric vector for
reaction $R_j$.

The model can be represented as the following expression involving
$N_r$ different unit rate Poisson
processes\cite{anderson2011continuous} $Y_j$, given by:
\begin{equation}\label{eq:RTC}
X(t) = X(0) + \sum_{j=1}^{N_r} \nu_j Y_j \left (\int_0^t
  \alpha_j(X(s)) \right ) ds.
\end{equation}

The master equation for this model is only solvable in certain
situations, for example monomolecular networks\cite{jahnke2007solving}, or for
steady-state solutions for weakly reversible deficiency
zero networks\cite{anderson2010product,anderson2016product}. Trajectories for this system can be
sampled exactly, for instance using the Gillespie SSA\cite{gillespie1977exact}, or
its
variants\cite{gillespie2007stochastic,cao2004efficient,anderson2007modified}.




\subsection{Likelihoods arising from multiscale stochastic reaction networks}
Suppose that we are able to exactly observe the number of molecules of
each chemical species in a system that can be well approximated by
\eqref{eq:RTC}. {Each reaction in the system involves the firing
  of a Poisson process with rate which is a function of the number of
  reactant molecules present, and the reaction rate constants. The likelihood
  of each reaction is simply the product of the probability density of
  an exponential random
  variable which gives the waiting time since the last reaction, with
  the mass function of a
  multinomial distribution describing the likelihood of each type of
  reaction from that state. Since the
  reactions are assumed to be independent, the likelihood of the whole
  path is a product of all the individual reaction likelihoods. This
  can be computed in a more efficient way by considering all reactions
  in the data from the same point in the state space at the same
  time. A detailed derivation of this is given in the
  supplementary material.

If we are only able to observe the slow species, then we cannot write
down an explicit likelihood for the path. One approach is to try to
simultaneously infer the
missing data of the fast variables, but this is computationally
intractable in all but the simplest of cases. The other option is to
invoke a multiscale approximation to the dynamics. This allows us to
approximate effective propensities for the slow reactions, and hence
to approximate the likelihood of paths of the slow variables. A
detailed derivation of this likelihood is also given in the
supplementary material.

In this work, we will consider two different multiscale
approximations. The first is the quasi-equilibrium approximation
which makes the assumption that fast reactions
enter quasi-equilibrium on a timescale which is negligible with
respect to the timescale on which the slow dynamics in the system are
evolving. This approximation is the basis of several
methods\cite{weinan2007nested,cao2005slow}. This approach can work well when there is a large timescale gap
between the fast and slow reactions in a system, but where the
timescale gap is less pronounced, it can lead to significant
errors\cite{cotter2016error}. Another approach is the
constrained multiscale algorithm (CMA)\cite{cotter2011constrained,cotter2016constrained}, based in part on the
equation-free approach to multiscale computations\cite{kevrekidis2003equation,erban2006gene}. This approach also assumes quasi-equilibrium in the
system, but takes into account the effect that the slow reactions can
have on the invariant distribution of the fast variables in the
system. For a more detailed description of this method, please refer
to the literature\cite{cotter2016constrained}.

}
\section{Numerical Examples}\label{sec:num}
In this section we consider two examples of chemical systems to
demonstrate the effectiveness of transport ETAIS algorithms for
sampling from the
challenging posterior distribution on the reaction parameters that
arise. 



\subsection{A Multiscale Chemical System}\label{sec:chem_multiscale}

First we consider a simple multiscale example involving two chemical
species $S_1$ and $S_2$,
\begin{equation}\label{eqn:full_system}
	\emptyset \xrightarrow{k_1} S_1 \xleftrightarrows[k_3]{k_2} S_2 \xrightarrow{k_4} \emptyset.
\end{equation}
Each arrow represents a reaction from a reactant to a product, with
some rate constant $k_i$, and where mass action kinetics is
assumed. The parameters $k_i$ are non-negative, and $\mathbf{k} =
[k_1,\dots,k_4]^\top \in \mathbb{R}_+^4 = \mathcal{X}$. We denote the
population count of
species $S_i$ by $X_i \in \mathbb{N}_0$. 
We consider this system with the following parameters:
\begin{equation}\label{eq:params1}
k_1V = 100, \qquad k_2 = k_3 = 10, \qquad k_4 = 1,
\end{equation} 
where $V$ is the volume of the reactor.
 In this 
parameter regime the reactions $R_2\colon S_1\rightarrow S_2$ and $R_3\colon S_2\rightarrow S_1$ occur
more frequently than the other reactions. Notice that both
chemical species are involved in fast reactions. However, the quantity
$S = X_1 + X_2$, is conserved by both of the fast reactions,
and as such, this is the slowly changing quantity in this system.

We observe this system over the time period $t \in [0,500]$ with
initial condition $S_1(0) = S_2(0) = 0$, but we assume that we are only able to observe
the slow variable $S(t) = S_1(t) + S_2(t)$. 
The effective dynamics of $S$ can be approximated by a system of only
two reactions:
\begin{equation}\label{eqn:QSSA_system}
	\emptyset \xrightarrow{\alpha_1(S)} S \xrightarrow{\bar{\alpha}_4(S)} \emptyset,
\end{equation}
where here the propensities are shown as opposed to the rates.

The effective propensity $\bar{\alpha}_4(S)$ can be  approximated through
application of a multiscale approximation, as detailed in {the
  supplementary material}, and in more detail in \cite{cotter2016constrained}. These effective
propensities can often be computed without the need for expensive
simulations, in particular when the fast subsystem is deficiency zero,
which is often the case\cite{anderson2010product,anderson2016product}. For this very
simple example, the dependence of these effective
propensities on the original rate parameters can be explicitly
understood. Under the QEA,
the effective rate of degradation of the slow variable $S$ is given by
\begin{equation}
	\bar{\alpha}_4^{\text{QEA}}(s) = \frac{k_2k_4s}{k_2+k_3}.
\end{equation}
Similarly, the analysis of the constrained system as discussed in~\cite{cotter2016constrained} yields the effective propensity
\begin{equation}\label{eqn:chem_CMA_rate}
	\bar{\alpha}_4^{\text{CMA}}(s) = \mathbb{E}_{\text{CMA}}\left[k_4X_2|S=s\right] = \frac{k_2k_4s}{k_2+k_3+k_4}.
\end{equation}

Our observations are uninformative about the reaction rates $k_2,
k_3$, and $k_4$, as for each multiscale approximation there
is a manifold $\mathcal{M} \subset \mathcal{X}$ along which the effective propensity
$\hat{\alpha}_4$ is invariant, leading to a highly ill-posed inverse
problem. This type of problem is notoriously difficult to
sample from using standard MH algorithms, as the algorithms quickly
gravitate towards the manifold $\mathcal{M}$ but then
exploration around $\mathcal{M}$ is slow.

We aim to recover three different posterior distributions. In the
first, we assume that we can fully and exactly observe the whole state
vector for the whole of the observation time window $t \in [0,T]$. In
the second and
third, the same data are used, but projected down to one dimensional observations of the slow
variable $S = X_1 + X_2$. In these last two examples, we approximate the
posterior using QEA and CMA approximations of the
effective dynamics respectively.

In all three cases, we find the posterior distribution for $\mathbf{k}
\in \mathcal{X} = \mathbb{R}_+^4$. These four parameters are assigned Gamma prior distributions,
\begin{equation}
	k_i \sim \text{Gamma}(\cdot; \alpha_i, \beta_i), \quad \text{for} \quad i = 1, \dots, 4.
\end{equation}
The hyper-parameters corresponding to each of these prior distributions are given in Table~\ref{tab:multiscale_priors}. These priors are the same for each of the three posterior distributions.

\begin{table}
\centering

\begin{tabular}{lrrrr} 
	\toprule
	Parameter$\backslash$Dimension $i$ & 1 & 2 & 3 & 4  \\ \cmidrule(lr){1-5} 
	$\alpha_i$ & 150 & 5 & 5 & 3 \\
	$\beta_i$ & $\frac{15}{9}$ & $\frac{5}{12}$ & $\frac{5}{12}$ & $1$ \\ \bottomrule 
\end{tabular}
\caption{Hyper-parameters in the prior distributions for the multiscale problem described in Section~\ref{sec:chem_multiscale}.}
\label{tab:multiscale_priors}
\end{table}

\begin{figure}[htb]
\centering
\includegraphics[width=0.9\textwidth]{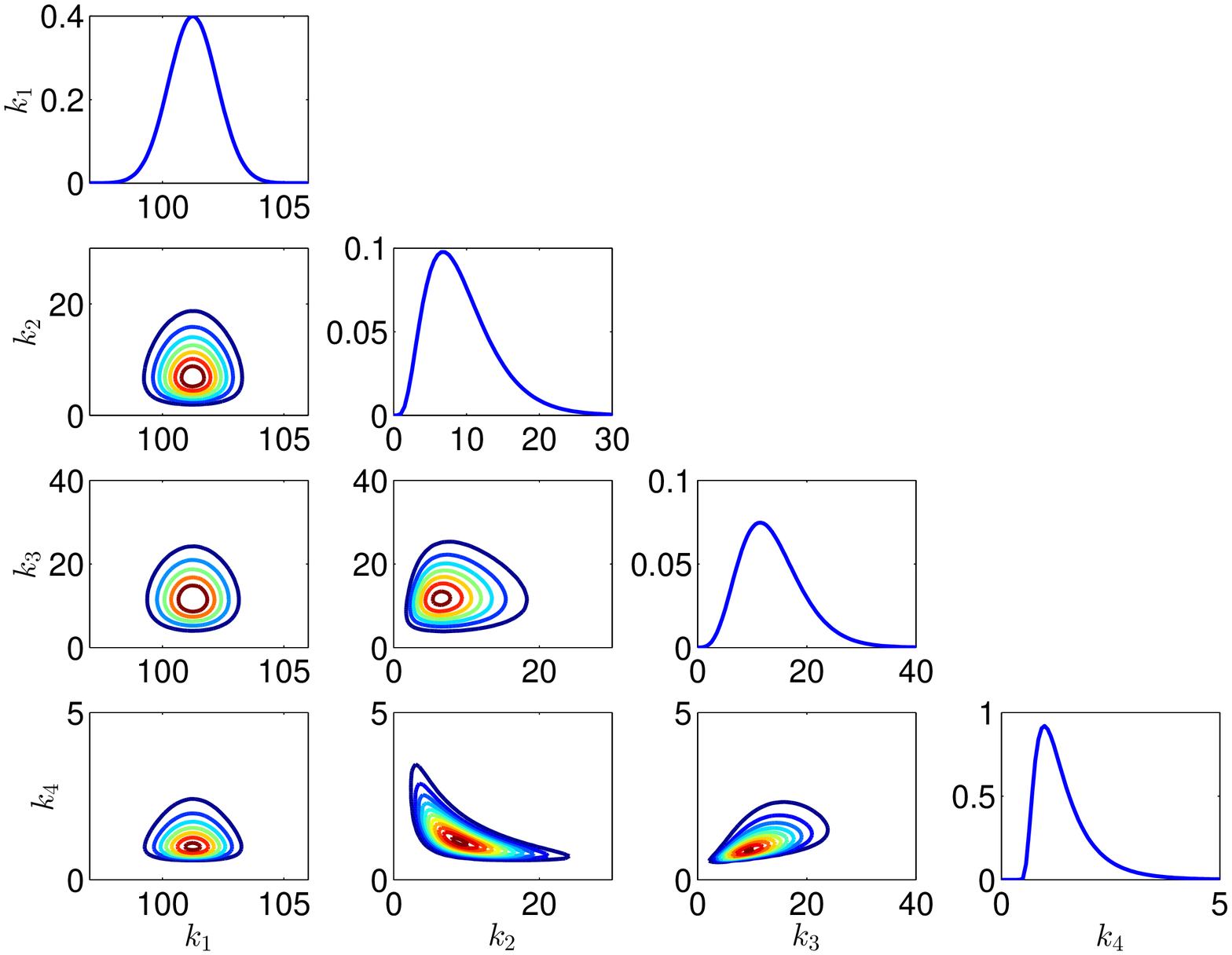}
\caption{CMA approximation of the posterior arising from observations of
  the slow variable $S = X_1 + X_2$ for system
  \eqref{eqn:full_system}  with true rates given by \eqref{eq:params1}.}
\label{fig:chem_CMA_posterior}
\end{figure}

Figure~\ref{fig:chem_CMA_posterior} shows how this posterior looks
when we use the CMA to model the effective degradation propensity
$\bar{\alpha}_4$. Although this posterior does not appear to be
impenetrably difficult to sample using MCMC from these plots, it most certainly is. The posterior
density is concentrated close to a curved manifold. Since the manifold
is curved, the joint marginals do not look unreasonable. However,
the distribution
is highly peaked around this curved manifold, making it very difficult
to sample from efficiently.

We consider several different algorithms for sampling from these
distributions. First we implement both the ETAIS and MH algorithms with
a Gaussian proposal distribution. In the case of the ETAIS algorithm,
this is a Gaussian mixture proposal distribution, and in the MH
algorithm this is equivalent to a ``vanilla'' random walk. The proposal
distribution uses a covariance matrix which has been constructed using
the sample covariance of a sample produced using a standard MH-RW
algorithm. This is necessary since these algorithms otherwise fail to
converge in a reasonable time for
this problem. We also compare the ETAIS and MH algorithms when using a
transport map proposal distribution, as was discussed in detail in Chapter~\ref{sec:TETAIS}.

In practice we cannot ensure that the approximate map $\tilde{T}$
is uniquely invertible over the whole of $R$ and so $\tilde{T}$ is not
truly bijective. This leads to problems for our strictly positive
state space, $\mathcal{X} \subset \mathbb{R}_+^d$, and so proposals in
$R$ do not necessarily map back onto $\mathcal{X}$. {This can be
  avoided, however, by employing a logarithmic preconditioner for the
  transport map. The proposal distributions are labelled with a T for transport map and when using the intermediate space we prepend `log' to the proposal method, e.g. MH-logRW and MH-logTRW, with, ETAIS-logRW and ETAIS-logTRW.}



{ The inclusion of the composition with the logarithmic map} means that we must alter our importance weight definition to reflect the pullback from $R$ through $\mathcal{Y}$. The weight is now
\begin{equation}
	w_i(\theta) = \frac{\pi(\theta|\mathbf{R},\mathbf{T})}{\chi(\tilde{T}\circ\log(\theta)|\tilde{T}\circ\log(\theta^{(i-1)}))|J_{\tilde{T}\circ\log}(\theta)|},
\end{equation}
where $\theta$ is a proposed new state, $\theta^{(i-1)}$ is the ensemble of states from the previous iteration, and $J_{\tilde{T}\circ\log}(\theta)$ is the Jacobian of the composition of the two maps. This Jacobian is straightforward to calculate,
\begin{equation}
	|J_{\tilde{T}\circ\log}(\theta')| = |J_{\tilde{T}}(\log(\theta'))||J_{\log}(\theta')|,
\end{equation}
where the first determinant is computed as in
\eqref{eqn:separable_jacobian}, and the second is given by
\begin{equation}
	|J_{\log}(\theta')| = \prod\limits_{i=1}^d \frac{1}{\theta'_i}.
\end{equation}

For this problem, we continue to use monomials in each dimension in our transport map construction. We use polynomials of total order $p=3$ as the basis functions, i.e.
\begin{equation}
	T_i(\theta) = \sum_{\mathbf{j}\in\mathcal{J}^{\text{TO}}_i(p)} \gamma_{i,\mathbf{j}}\psi_{\mathbf{j}}(\theta) \quad \text{where} \quad \psi_\mathbf{j}(\theta) = \prod\limits_{k=1}^i \theta_k^{j_k},
\end{equation}
and
\begin{equation}
	\mathcal{J}^{\text{TO}}_i(p) = \{\mathbf{j} \in \mathbb{N}^d_0\ |\ \|\mathbf{j}\|_1 \leq p, \ \text{and}\ j_k = 0\ \forall k > i\}.
\end{equation}

We will use the MT\cite{russ2017parallel} resampler with an ensemble size of
$M=500$. This increase in ensemble size in comparison with the previous
example in Section \ref{sec:conv} is to allow for the increase in parameter dimension.

To measure the convergence of the sampling methods in this section, we
will compare the convergence of the mean of each parameter. We
approximate $\mathbb{E}(\mathbf{k})$ using 2.4 billion samples from
the MH-RW algorithm. We do this for each of the eight algorithms we
have so far discussed. Convergence is shown only for the CMA
approximation of the posterior with effective rate for
the degradation of $S$ given by \eqref{eqn:chem_CMA_rate}, but we
expect very similar results for the other multiscale approximation of
posterior distributions
discussed.

{
\begin{table}[!h]
	\centering
	\begin{tabular}{lrrrr}
	\toprule
		 & \multicolumn{2}{c}{MH} & \multicolumn{2}{c}{ETAIS} \\ \cmidrule(lr){2-3} \cmidrule(lr){4-5}
		& logRW & log-TRW & logRW & log-TRW \\ \midrule
		$\beta_{\%}^*$	 	& 1.7e-1 &2.7e-2 & - & - \\
		$\beta_{\text{ESS}}^*$	       & -            & -          & 1.2e-0 & 1.5e-1 \\
		ESS/$M$		 	 & -            & -            & 6.0e-2 & 3.5e-1  \\
	\bottomrule
	\end{tabular}
	\caption{Optimal scaling parameters for the MH and ETAIS
          algorithms applied to the constrained multiscale problem in
          Section~\ref{sec:chem_multiscale}. MH parameters optimised
          by acceptance rate, and ETAIS parameters optimised using
          effective sample size. ESS per sample also shown for the
          ETAIS algorithms.}
\label{tab:chem_multiscale_scaling}
\end{table}}


The optimal scaling parameters for the proposal distributions are
given in Table~\ref{tab:chem_multiscale_scaling}. These are
precomputed by running the algorithms for different parameter
values. MH-RW based algorithms are tuned to an acceptance rate close
to the optimal $23.4\%$, and ETAIS algorithms are tuned to maximise the
effective sample size (ESS). The optimal parameter values for the MH
algorithms are not particularly informative about their performance,
since they relate to algorithms operating on different
transformations. {The ESS per sample is also displayed in
Table~\ref{tab:chem_multiscale_scaling} for the ETAIS implementations,
and is highest for the logTRW variant
of the ETAIS algorithm, as we would hope.} We note here that, as
outlined in \cite{cotter2015parallel}, it is possible to construct
adaptive versions of ETAIS which automatically tune these algorithmic
parameters, but we compare against the static optimal parameter values
for all algorithms to ensure fairness.

\begin{figure}[!htb]
\centering
\includegraphics[width=0.495\textwidth]{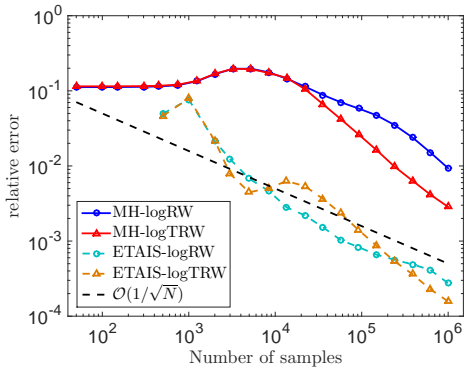}
\caption{Convergence of four different MCMC algorithms for the CMA
  approximation of the  posterior arising from the example in Section
  \ref{sec:chem_multiscale}. {Plots show average
    convergence over 32 repeats.}}
\label{fig:chem_multiscale_L2}
\end{figure}

{
Convergence of the four algorithms for this example are shown in
Figure~\ref{fig:chem_multiscale_L2}. We first note the poor
performance of the MH based algorithms, despite being preconditioned
with a pre-computed sample covariance. Only the MH-TRW is at all competitive with the ETAIS algorithms. During the simulation interval, the MH-TRW algorithm has not settled down to the expected $\mathcal{O}(1/\sqrt{N})$ rate which means that the estimate is still biased by the long burn-in time. As we have seen in previous examples, the burn-in time for the ETAIS algorithm is negligible.

This proposal method
becomes more stable as we increase either the ensemble size, or the
number of iterations between updates of the transport map,
$T$. Overall we see the smallest Monte Carlo errors for a given number
of likelihood evaluations coming from the ETAIS-logTRW algorithm. This
is due to weight collapse in some of the 32 repeats of the ETAIS-logRW
algorithm, demonstrating that the transport map stabilises the ETAIS methodology.
}

\begin{figure}[!htb]
\centering
\subfigure[Pushforward of the posterior sample through $\hat{T}$ with log preconditioner.]{\includegraphics[width=0.8\textwidth]{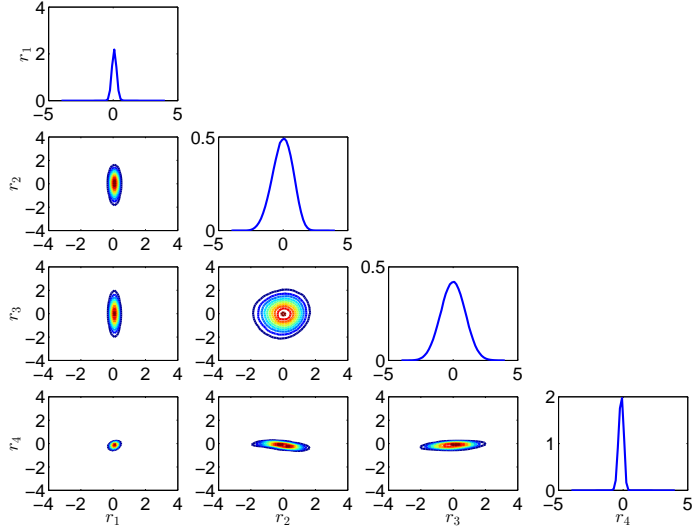}}
\caption{{Pushforward of the target measure on reference space for the CMA
  approximation of the  posterior arising from the example in Section \ref{sec:chem_multiscale}, found using the logTRW proposal distributions. Components are linked by the relation $r_i = T_i\circ\log(k_i)$.}}
\label{fig:chem_reference_spaces}
\end{figure}

{
We now look at the proposal distributions of the transport map accelerated algorithm. In Figure~\ref{fig:chem_reference_spaces}, we see the reference space found by mapping the posterior through $T\circ\log$.} For the most part, each of these marginal distributions can be recognised as a Gaussian. However, with the exception of $\mathbb{P}(r_2,r_3)$, we would not consider them to be close to a standardised $\mathcal{N}(0, \text{I})$ distribution. Before thinking that the transport map has not helped us to find a `nicer' space on which to propose new values we should consider that the dimensions are now (1) largely uncorrelated, and (2) the variances in each dimension are much more similar than they are in Figure~\ref{fig:chem_CMA_posterior}.

{We see that
$\text{var}(r_1)$ and $\text{var}(r_4)$ are much smaller than
$\text{var}(r_2)$ and $\text{var}(r_3)$. To combat this we have a
number of choices.} We might wish to learn the covariance structure of
the pushforward of the target onto the reference space, and
incorporate this information into the covariances of the mixture
components in an adaptive scheme. Another option is to increase the
total order of our index set. For these numerics we have chosen $p=3$,
but we can obtain reference spaces which are closer to
$\mathcal{N}_d(0, \text{I})$ by choosing a larger $p$.

{At this point, we note that if the transport map found is equal
  or close to the Knothe-Rosenblatt rearrangement, then the most
  efficient sampling strategy is clearly to independently sample from the
  reference measure and pull these samples back through the map onto
  (or close to) the target. However we see from Figure
  \ref{fig:chem_reference_spaces} that in practice for very complex
  targets, the approximation of the transport map is not as good, and
  as such the pushforward of the target onto the reference
  space differs greatly from the intended reference measure. In
  this case, the pullback of the reference measure will be far from
  the target distribution, and this independent sampling strategy
  will be inefficient. This motivates the use of the ETAIS algorithm
  to efficiently sample from the more complex distributions arising
  from the pushforward of the target distribution. 
  }

\subsubsection{Comparison of the Constrained and QEA approaches}

The convergence analysis in the previous section was performed for the constrained
approximation of the posterior distribution. We now look at the
differences in the CMA and QEA approximations of the posterior distribution
arising in this problem. Recall that the approaches differed only in the form of the effective degradation propensity $\hat{\alpha}_4$,
\begin{equation}
	\hat{\alpha}_4^{\text{QEA}}(s) = \frac{k_2k_4s}{k_2+k_3} \quad \text{and} \quad \hat{\alpha}_4^{\text{CMA}}(s) = \frac{k_2k_4s}{k_2+k_3+k_4}.
\end{equation}
This difference in the denominator causes a shift in the manifold on
which the posterior is concentrated.



In the QEA approximation, the effective rate of degradation of $S$ is
given by $\hat{k}_4^{\text{QEA}} = k_2k_4/(k_2+k_3)$, and as such, our
observations should be informative about this quantity. Using the CMA,
this effective rate is approximated by $\hat{k}_4^{\text{CMA}} =
k_2k_4/(k_2+k_3+k_4)$. To validate our inferences on $k_2$, $k_3$ and
$k_4$ we would like to discover which model is most accurate and
informative about these quantities, and which model gives us results
which are most similar to that which can be obtained from the full
model, with all reactions (including fast reactions) observed.

A conventional way to compare two models under a Bayesian framework is
to calculate the Bayes factors~\cite{chen2012monte}. The marginal densities for the data, evaluated at the observed data, given the models are displayed in Table~\ref{tab:chem_Bayes_marginals}.

\begin{table}[!htb]
\centering
\begin{tabular}{cccc}
	\toprule
	Model description & $k$ & \quad $\mathbb{P}(D =
                                  (\mathbf{R},\mathbf{T})^\top|\mathcal{M}_k)$
  & Bayes' factor  \\ \cmidrule(lr){1-4}
	Full & 0 & 6.8e-3 & 1 \\
	CMA & 1 & 3.2e-3 & 2.09 \\
	QEA & 2 & 1.7e-3 & 4.10 \\ \bottomrule
\end{tabular}
\caption{Marginal distributions and Bayes factors for the data
  $(\mathbf{R},\mathbf{T})^\top$ for each model considered in
  Section~\ref{sec:chem_multiscale}. The Bayes' factors are computed
  with respect to the ``full'' model.}
\label{tab:chem_Bayes_marginals}
\end{table}

Computing the Bayes factors from Table~\ref{tab:chem_Bayes_marginals}
we see that we should of course prefer the model { in
which we observe all reactions perfectly and are able to explicitly
compute the likelihood}. However this model is not
significantly better than the CMA model ($B_{0,1} = 2.09 < 3.2$,
~\cite{kass1995bayes}). The Bayes factor $B_{0,2} = 4.1 > 3.2$ tells
us that we should significantly prefer the full model to the QEA
approximation of the posterior. These Bayes factors provide evidence
for our
argument that the constrained approximation of the posterior is
superior to the QEA approximation. 

\begin{figure}[!htb]
\centering
\subfigure[Marginal for $\hat{k}_4^{\text{QEA}}$.]{\includegraphics[width=0.495\textwidth]{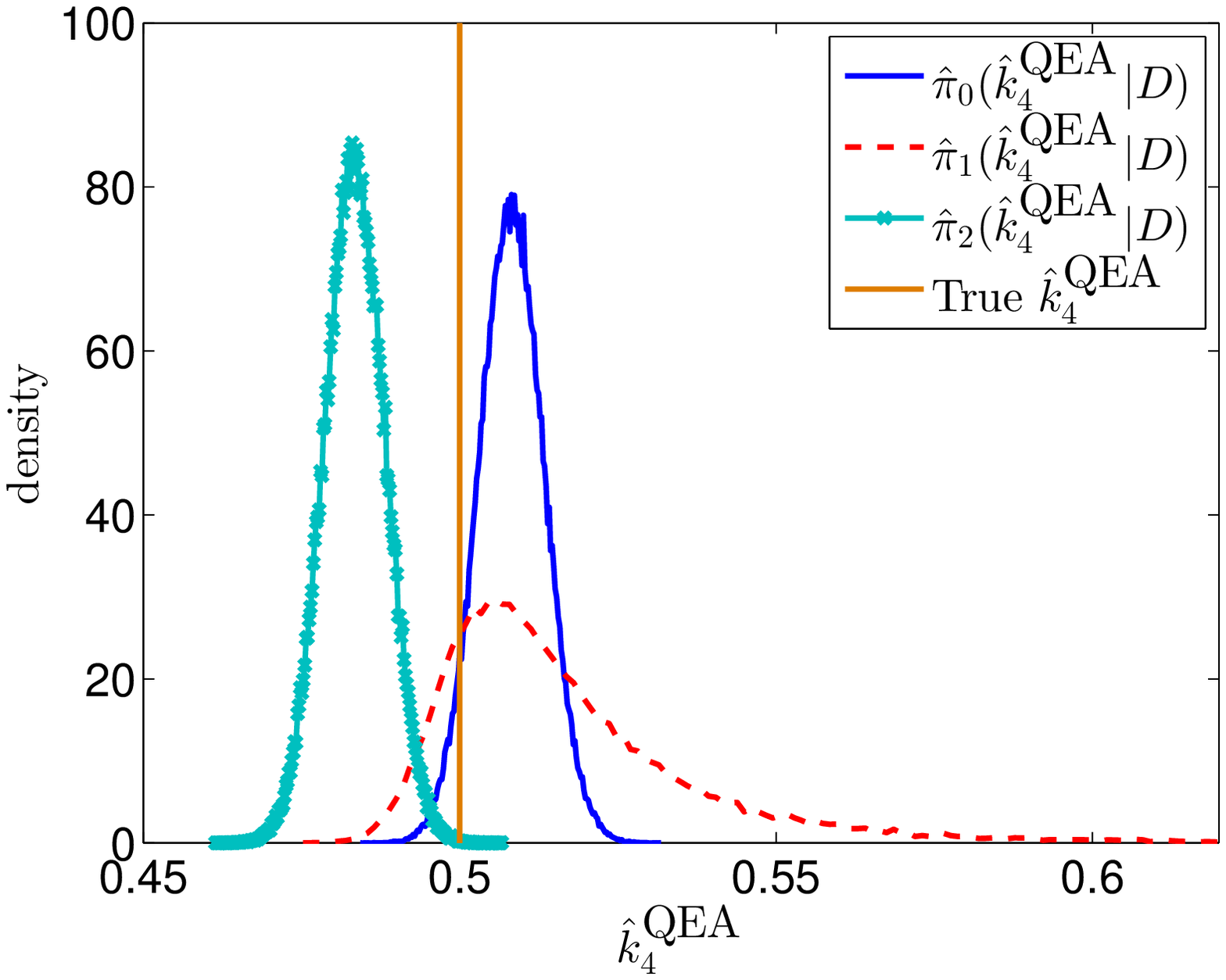}}
\subfigure[Marginal for $\hat{k}_4^{\text{CMA}}$.]{\includegraphics[width=0.495\textwidth]{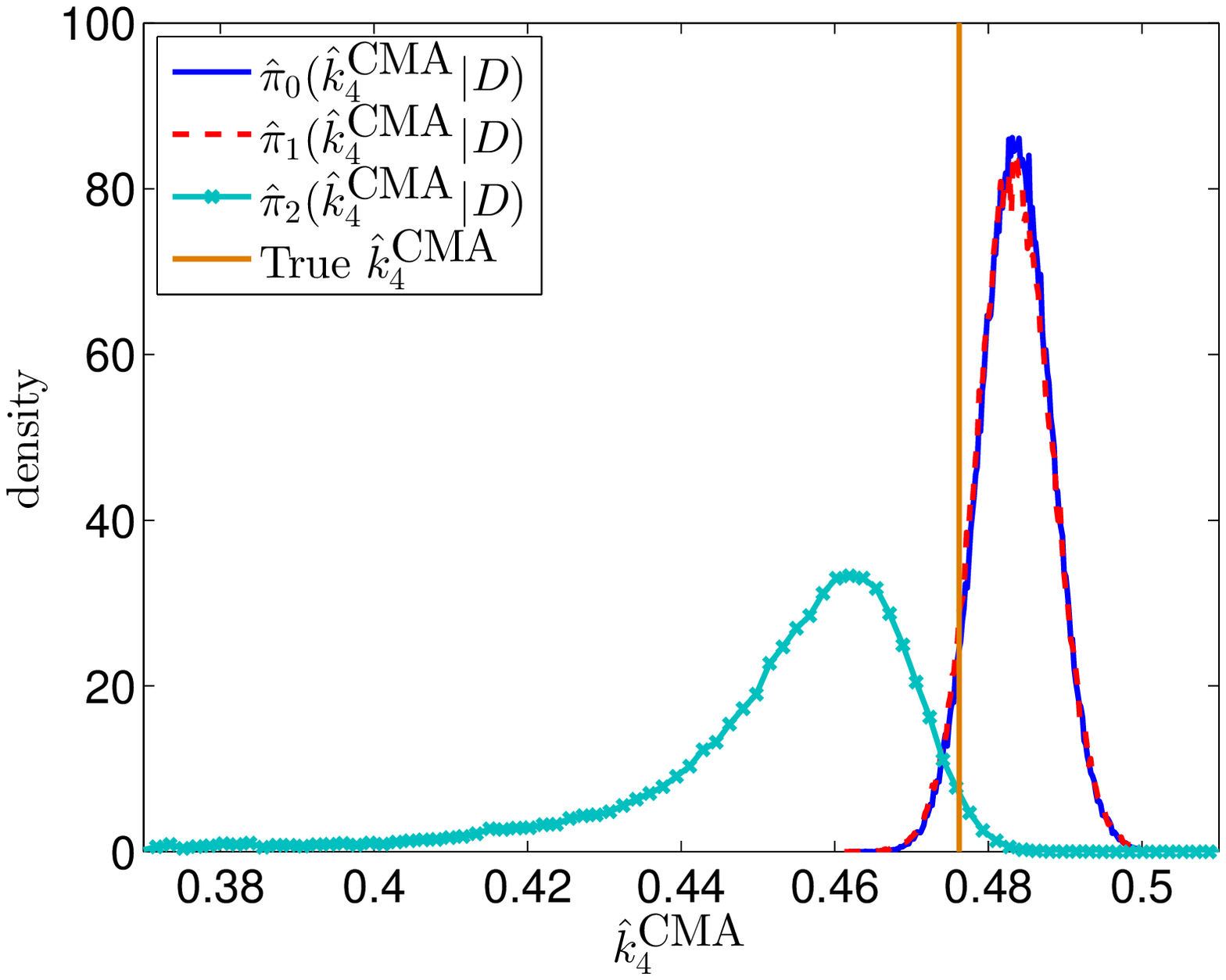}}
\caption{Comparison of the approximate marginal densities for the
  quantities $\hat{k}_4^{\text{QEA}}$ and $\hat{k}_4^{\text{CMA}}$
  across the three posterior densities $\pi_i = \mathbb{P}(\theta_i|D,\mathcal{M}_i)$ for $i = 0, 1, 2$.}
\label{fig:chem_model_comp}
\end{figure}

Figure~\ref{fig:chem_model_comp} displays more compelling evidence in
this direction. Here we show the marginal distributions for the
quantities $\hat{k}_4^{\text{QEA}}$ and $\hat{k}_4^{\text{CMA}}$ with
respect to each of the three posterior distributions, $\pi_i(\cdot|D,
\mathcal{M}_i)$, $i = 0, 1, 2$. Firstly, in
Figure~\ref{fig:chem_model_comp} (a), displaying the marginals for
$\hat{k}_4^{\text{QEA}}$ we see that $\pi_0$, the density arising from
the full model with all reactions observed, is peaked a distance away
from $\pi_2$, QEA approximation. The CMA approximation $\pi_1$, in contrast, is
peaked close to peak of $\pi_0$, but there is more uncertainty here
since we do not observe the fast reactions.

In contrast, in Figure~\ref{fig:chem_model_comp} (b), the marginal
densities for $\hat{k}_4^{\text{CMA}}$ for $\pi_0$ and $\pi_1$ are
barely distinguishable, even with the vastly reduced data available in
the posterior $\pi_1$ with CMA approximation. The QEA approximation
$\pi_2$ is again peaked far away from both of the other distributions.
This demonstrates that the data regarding the slow variable
$S$ is only informative about $k_1 $ and $\hat{k}_4^{\text{CMA}}$, as
predicted by the constrained multiscale approximation.

\subsection{Gene Regulatory Network}\label{sec:grn}
In this example, we look at a model of a gene regulatory network (GRN)
mechanism~\cite{kaern2005stochasticity,guido2006bottom,becskei2000engineering}
used by a cell to regulate the amount of a protein, $P$,
present. Proteins can bind in pairs to form a dimer $D$. The dimer can
bind to the gene $G$ and convert it into a ``switched off'' state $G'$, in which the
production of mRNA molecules $M$ is inhibited. The mRNA encodes for
production of the protein $P$, and both $P$ and $M$ can degrade. 
The eight reactions are given in Equation~\eqref{eqn:GRN_reactions}.
\begin{align}
	P+P \xleftrightarrows[k_2]{k_1} D \quad & \quad G+D \xleftrightarrows[k_4]{k_3} G' \nonumber\\
	G \xrightarrow{k_5} G+M \quad & \quad M\xrightarrow{k_6} M+P  \label{eqn:GRN_reactions} \\
	P \xrightarrow{k_7} \emptyset \quad & \quad M \xrightarrow{k_8} \emptyset \nonumber
\end{align}

We create trajectories of this system, using the Gillespie SSA, with the following parameters:
\begin{eqnarray}
k_1 = 0.04, \qquad
k_2 = 5000, \qquad
k_3 = 100, \qquad
k_4 = 1, \\ 
k_5 = 0.5, \qquad
k_6 = 2, \qquad
k_7 = 0.2, \qquad
k_8 = 0.05,
\end{eqnarray}
over the time period $t \in [0,500]$.

T-cells can be manufactured which bind to
proteins of a certain type, and cause them to phosphoresce, allowing
for approximate observations of the concentration levels of that
protein. However, the T-cells will not necessarily differentiate
between monomers and dimers of the protein. We approximately replicate such a
scenario, in a zero-observational noise setting, by assuming that we
can only observe $T=P + 2D + 2G'$, the total number of protein
molecules in the system, alongside $M$, the concentration of mRNA. The
number of dimers, and the status of the genetic switch, are assumed
unobservable. In
order to sample from the posterior distribution on the reaction
parameters in the system given these observations, we can either integrate over all possible
trajectories of the fast variables (which is computationally
intractable), or as in Section \ref{sec:chem_multiscale} we can use a
multiscale method to approximate the posterior. 

The effective dynamics of $T$ and $M$ are then given by:
\begin{align}
	\emptyset \xrightarrow{\hat{\alpha}_5} M \quad & \quad M\xrightarrow{\alpha_6} M+T  \label{eqn:GRN_eff} \\
	T \xrightarrow{\hat{\alpha}_7} \emptyset \quad & \quad M \xrightarrow{\alpha_8} \emptyset \nonumber
\end{align}
Here, the propensities are displayed, as opposed to the rates, with
approximations of the propensities $\hat{\alpha}_5$ and
$\hat{\alpha}_7$ to be computed. We omit the derivation of these
effective propensities, using the CMA methodology, for brevity, but
interested readers can see more details in \cite{russ2017parallel}.

\subsubsection{Target Distribution}
As alluded to in the previous section, we aim to sample from the CMA
approximation of the posterior distribution. This example is more involved
than the previous one, since we have an eight dimensional parameter
space corresponding to the eight reaction rates in the system
\eqref{eqn:GRN_reactions}. Half of those reactions are
unobservable, and their effect on the posterior is only felt through
their effect on the effective propensities $\hat{\alpha}_5$ and
$\hat{\alpha}_7$, and observations of the frequencies of these reactions.

The priors for this problem were chosen to be fairly uninformative,
and a list of the hyper-parameters corresponding to each Gamma prior can be found in Table~\ref{tab:grn_priors}.

\begin{table}[!h]
\centering
\begin{tabular}{lrrrrrrrr}
	\hline
	Dimension $i$ & 1 & 2& 3& 4 & 5 & 6 & 7 & 8 \\ \hline
	$\alpha_i$ & 2 & 100 & 100 & 3 & 3 & 3 & 2 & 2\\ \hline
	$\beta_i$ &50 & 0.02 & 1 & 1 & 0.6 & 1 & 50 & 50 \\ \hline
\end{tabular}
\caption{Hyper parameters for the Gamma priors on each of the reaction rates in the GRN example in Section~\ref{sec:grn}.}
\label{tab:grn_priors}
\end{table}

\begin{figure}[htb]
\centering
\makebox[\textwidth][c]{\includegraphics[width=1.2\textwidth]{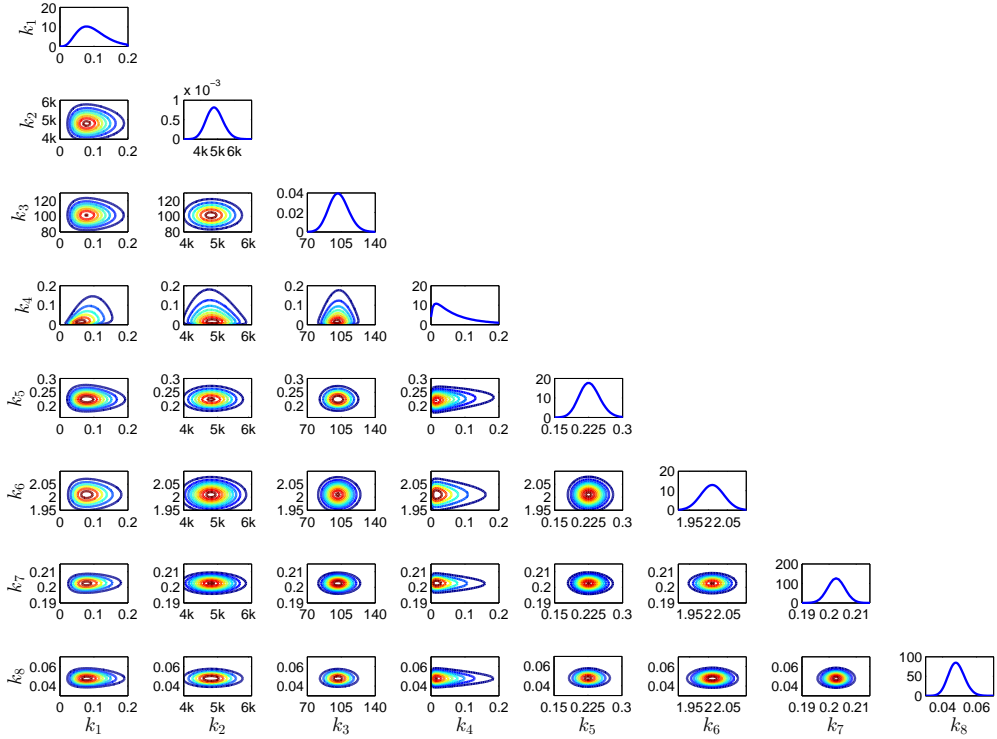}}%
\caption{Marginal densities of the CMA approximation of the posterior
  distribution for the GRN example in Section~\ref{sec:grn}.}
\label{fig:GRN_posterior}
\end{figure}

The one- and two-dimensional marginal distributions for the full
posterior distribution are displayed in
Figure~\ref{fig:GRN_posterior}. The posterior does not appear from
these joint marginals to contain many
interesting correlation structures, although several dimensions have
leptokurtic tails, which are difficult for the standard ETAIS algorithm
to sample from. However as with the previous example, the density is
concentrated close to a lower-dimensional curved manifold. The marginal densities also vary over very different
scales, which could lead to problems with the algorithms which do not
benefit from the transport map.

\subsubsection{Implementation}

We apply the {four logRW} proposal distributions, with and
without transport map acceleration, which were discussed in
Section~\ref{sec:chem_multiscale} to this posterior. As with this
example, the log conditioner is required so that our transport map is a function $T\colon\mathbb{R}^d\rightarrow\mathbb{R}^d$, rather than a map from $\mathbb{R}_+^d$ to $\mathbb{R}^d$.

The transport map setup has not changed from the previous
example. However, due to the scaling of the weights we have had to be
more careful about which samples we use to build our map. We require
that the objective function $C(T)$, from
Equation~\eqref{eqn:TETAIS_objective} is convex, which requires that
the second derivative, given in Equation~\eqref{eqn:TETAIS_hessian},
is positive definite. This positive definite property depends on all
weights in our sample being strictly positive, which is not always
possible on a computer, where very small positive numbers can be
rounded down to absolute zero. For this reason we filter out any samples from the optimisation problem where the weight is a numeric zero. This does not affect the validity of the map since a weight zero would not contribute to the expectation.

In all {four} algorithms we adapt the proposal variances during the
burn-in phase of the algorithms to avoid the need for a huge ensemble size to approximate the posterior with a mixture of isotropic Gaussian kernels.
We use the MT resampler, this time with an ensemble size of $M=2500$. The increase in ensemble size compensates for the increase in dimension from four to eight.

As in the previous chemical reaction example, we measure convergence of the algorithms only through the convergence of the mean. We use the sample Mahalanobis distance, with `true' covariance and `true' mean built from a sample of size 2.4 billion using the MH-RW algorithm.

\subsubsection{Convergence results}

The scaling parameters used are selected by optimising the acceptance
rate for the MH algorithms, and optimising the effective sample size
for the ETAIS algorithms. The optimal parameters are given in
Table~\ref{tab:grn_scaling_parameters}. Here, for Metropolis-Hastings
variants, $\beta_{\%}^*$ is the variance of the proposal, optimised
for an acceptance rate of $50\%$. For the ETAIS variants,
$\beta_{\text{ESS}}^*$ is the variance of each of the proposal kernels
around each particle, optimised numerically for the maximal effective
sample size (ESS). ESS/$M$ denotes the ratio of ESS to the number of
particles, with values closer to one denoting a more efficient algorithm.

\begin{table}[!h]
	\centering
	\begin{tabular}{lrrrr}
	\toprule
		 & \multicolumn{2}{c}{MH} & \multicolumn{2}{c}{ETAIS} \\ \cmidrule(lr){2-3} \cmidrule(lr){4-5}
		& logRW & log-TRW & logRW & log-TRW \\ \midrule
		$\beta_{\%}^*$	 	& 3.9e-1 &6.0e-1 & - & - \\
		$\beta_{\text{ESS}}^*$	       & -            & -          & 1.3e-0 & 5.0e-1 \\
		ESS/$M$		 	 & -            & -            & 4.7e-2 & 1.6e-1  \\
	\bottomrule
	\end{tabular}
	\caption{Scaling parameters for the sampling algorithms applied to the GRN example in Section~\ref{sec:grn}.}
	\label{tab:grn_scaling_parameters}
\end{table}

From the effective sample sizes shown in
Table~\ref{tab:grn_scaling_parameters} we can see the improvement to
the efficiency of the algorithm both by transforming the parameter
space into something closer to Gaussian.

\begin{figure}[!htb]
\centering
\includegraphics[width=0.495\textwidth]{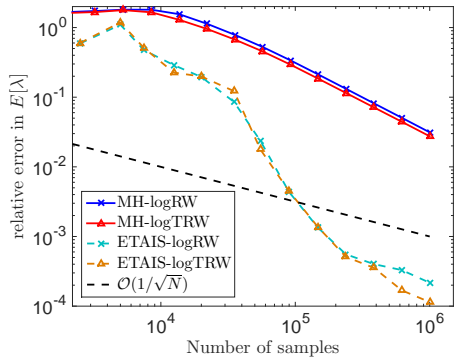}
\caption{Convergence of four different MCMC algorithms for the CMA
  approximation of the  posterior arising from the example in Section
  \ref{sec:grn}. {Plots show average
    convergence over 32 repeats.}}
\label{fig:grn_L2}
\end{figure}

Due to the numerical cost of calculating the full relative $L^2$ errors for this posterior, we quantify the convergence speeds of these algorithms using the sample Mahalanobis distance between the sample mean and the `true' mean. This convergence analysis is shown in Figure~\ref{fig:grn_L2}.

We believe that an ensemble size of 2500
is over-cautious, despite the larger number of dimensions in this
example, and that the required ensemble size to sample
from these posteriors is reduced by the use of the transport map.

The second obvious feature of these convergence plots is that the ETAIS
algorithms outperform the MH algorithms by a large margin - roughly a
reduction of two orders of magnitude in the relative error over the
same number of samples.


{The transport map version exhibits more
consistent and stable convergence, with some of the 32 repeats for the
ETAIS-logRW algorithm producing rare large weights which hinder convergence.}
The large increase in the effective sample size
observed between the ETAIS-logRW and ETAIS-logTRW algorithms is
converted into an estimate which is twice as accurate after 1 million
samples. This corroborates with our claim that the transport map can
produce stable ETAIS implementations with smaller ensemble sizes than without.

{A similar pattern is seen in the MH algorithms, where the MH-logTRW algorithm performs the best,
although only marginally. }

\section{Discussion}\label{sec:conc}
In this paper, we have investigated the use of transport maps to
accelerate and stabilise ensemble importance sampling schemes. We
considered the ETAIS methodology here, but the approach used is
applicable to all such methods. The idea is powerful since it allows
adaptive importance sampling schemes to be efficient and stable for
smaller ensemble sizes. This in turn suggests that the improvements in
performance that adaptive importance sampling schemes can deliver
could be made applicable to a bigger class of problems.

{If good transport maps can be found, such that the pushforward
  of the posterior is close to a reference Gaussian, then the most
  efficient approach is to draw independent samples pulled back
  through the map onto the target space. However, for complex
  posteriors, in particular the type that we have observed here where
  due to sloppiness in the model, the posterior is concentrated on a
  low-dimensional manifold, a good map is hard to find. The numerical
  approximation of the optimal map leads to a measure on the reference
  space which is far removed from the reference Gaussian, and requires
  more considered exploration. Adaptive importance sampling schemes
  such as ETAIS are able to sample from such a measure
  more efficiently than Metropolis-Hastings equivalents.}

One issue that we have not looked at in depth, is that of the choice
of basis functions for the transport map, which is discussed in more
depth in \cite{parno2018transport}. It seems reasonable to
expect that this choice could have an impact on the quality of the
transport map, which in turn has an impact on the number of ensemble
members which are required to stably sample from the target
distribution. We explored some different combinations, including higher
order but sparse bases, but we did not observe a big difference in
performance. It would be an interesting avenue to explore how other
approaches, for example tailored, anisotropic, or even adaptively
chosen parameterisations, could enable the map to more accurately
capture the structure of the target density, reducing the required
ensemble size.

In particular we looked at the application of these methods to
problems arising in inverse problems for multiscale stochastic
chemical networks, where the posterior distributions which arise can
be highly concentrated along curved manifolds in parameter
space. Through these results we also demonstrated the improvement in
accuracy of inference for multiscale problems when more sophisticated
multiscale approximations are employed. Below we highlight two areas for future work.

\subsection{Sampling in higher
  dimensions}\label{sec:TETAIS_higher_dim}

One major issue with importance sampling-based algorithms is the curse of
dimensionality. This curse is two-fold, since as the target
distribution's dimension is increased, a larger ensemble size is often
required in order to capture the addition complexity of the density
that the higher dimension enables. Furthermore, the cost of the
resampling step within the algorithm grows with the ensemble size. For
example, the ETPF algorithm is $\mathcal{O}(n^2)$. However, the
simplification that the transport map makes has some potential to go
some way to tackling this problem.

Here, we will briefly discuss how transport maps could aid
with making moderately higher dimensional problems accessible to this
family of methods. It is important to emphasise here that these
methods are unlikely to be suitable in anything other than low
dimensional problems, but the idea that we present here, which is the
subject of future investigation, could make them applicable to more
problems which have more than a handful of parameters.

Armed with a high quality transport map, we can transform the target
distribution close to a Gaussian reference measure, which is
approximately uncorrelated in all dimensions. This lack of correlations allows us to
resample in each dimension separately. Resampling in a
single dimension allows for optimisations in resampling code, and also
means that the resampler is not affected by the curse of
dimensionality. 

In one
dimension, the ETPF algorithm can be implemented very efficiently. As
described in~\cite{reich2013nonparametric}, the coupling matrix has
all non-zero entries in a staircase pattern when the state space is
ordered. We can exploit this knowledge to produce
Algorithm~\ref{alg:ETPF_1d}, which is much faster than using the
simplex algorithm to minimise the associated cost function, and faster
than the MT algorithm\cite{cotter2015parallel}, which is itself a
greedy approximation to the output of the ETPF resampler. Moreover, the task of
running this resampler can then be parallelised over $d$
processors. This could be an area for further study.

\begin{table}[!htpb]
\begin{algorithm}[H]
\DontPrintSemicolon
\BlankLine
Sort the states, $\{(w_i, x_i)\}_{i=1}^M$, into ascending order.\;
Normalise the weights $p_i = w_i/\|w\|_1$.\;
Set $y_i \leftarrow 0$ for all $i=1,\dots,M$.\;
Set $c \leftarrow 0$\;
\For{$i \leftarrow 1, \dots, M$}{
	Set $t \leftarrow p_i$\;
	\While{$j \leq M$ and $t > 0$}
	{
		Set $s \leftarrow \left(M^{-1}-c\right) \wedge t$\;
		Increase $y_j$ by $M\times s\times x_i$.\;
		Decrease $t$ by $s$.\;
		Increase $c$ by $s$.\;
		\If{$t>0$}
		{
			Increase $j$ by 1.\;
			Set $c \leftarrow 0$.
		}
	}
}
Return $y$.\;
\caption{ETPF algorithm in one dimension.\label{alg:ETPF_1d}}
\end{algorithm}
\end{table}


\subsection{Adaptivity of ensemble sizes}
At initialisation, our transport map approximation is equal to the
identity, and the algorithm is essentially the standard ETAIS
algorithm. In order to build a good transport map approximation, we
require a good sample from the posterior distribution. Since our aim
is to apply this methodology to challenging problems, this is itself a
challenge. However, in the limit of the number of ensemble members
going to infinity, this approach is consistent and stable. However,
large ensemble sizes come with large computational costs.

We propose that the transport map acceleration could be used in
conjunction with an adaptive ensemble size mechanism. First we can
initialise the algorithm with a large ensemble size, and the identity transport map. As
we explore the density, the transport map quickly improves, making it
possible to sample efficiently and stably with a smaller ensemble
size. The problem then amounts to finding a reliable method to
approximate the lowest stable ensemble size.


\bibliographystyle{siam}
\bibliography{bibliography}

\begin{thebibliography}{10}

\bibitem{anderson2007modified}
{\sc D.~Anderson}, {\em A modified next reaction method for simulating chemical
  systems with time dependent propensities and delays}, The Journal of chemical
  physics, 127 (2007), p.~214107.

\bibitem{anderson2016product}
{\sc D.~Anderson and S.~Cotter}, {\em Product-form stationary distributions for
  deficiency zero networks with non-mass action kinetics}, Bulletin of
  mathematical biology, 78 (2016), pp.~2390--2407.

\bibitem{anderson2010product}
{\sc D.~Anderson, G.~Craciun, and T.~Kurtz}, {\em Product-form stationary
  distributions for deficiency zero chemical reaction networks}, Bulletin of
  mathematical biology, 72 (2010), pp.~1947--1970.

\bibitem{anderson2011continuous}
{\sc D.~Anderson and T.~Kurtz}, {\em Continuous time markov chain models for
  chemical reaction networks}, in Design and analysis of biomolecular circuits,
  Springer, 2011, pp.~3--42.

\bibitem{apgar2010sloppy}
{\sc J.~Apgar, D.~Witmer, F.~White, and B.~Tidor}, {\em Sloppy models,
  parameter uncertainty, and the role of experimental design}, Molecular
  BioSystems, 6 (2010), pp.~1890--1900.

\bibitem{becskei2000engineering}
{\sc A.~Becskei and L.~Serrano}, {\em Engineering stability in gene networks by
  autoregulation}, Nature, 405 (2000), p.~590.

\bibitem{bierkens2016zig}
{\sc J.~Bierkens, P.~Fearnhead, and G.~Roberts}, {\em The zig-zag process and
  super-efficient sampling for bayesian analysis of big data}, arXiv preprint
  arXiv:1607.03188,  (2016).

\bibitem{bogachev2005triangular}
{\sc V.~Bogachev, A.~Kolesnikov, and K.~Medvedev}, {\em Triangular
  transformations of measures}, Sbornik: Mathematics, 196 (2005), p.~309.

\bibitem{bouchard2018bouncy}
{\sc A.~Bouchard-C{\^o}t{\'e}, S.~Vollmer, and A.~Doucet}, {\em The bouncy
  particle sampler: A nonreversible rejection-free markov chain monte carlo
  method}, Journal of the American Statistical Association,  (2018), pp.~1--13.

\bibitem{bugallo2017adaptive}
{\sc M.~Bugallo, V.~Elvira, L.~Martino, D.~Luengo, J.~Miguez, and P.~Djuric},
  {\em Adaptive importance sampling: the past, the present, and the future},
  IEEE Signal Processing Magazine, 34 (2017), pp.~60--79.

\bibitem{cao2005slow}
{\sc Y.~Cao, D.~Gillespie, and L.~Petzold}, {\em The slow-scale stochastic
  simulation algorithm}, The Journal of chemical physics, 122 (2005),
  p.~014116.

\bibitem{cao2004efficient}
{\sc Y.~Cao, H.~Li, and L.~Petzold}, {\em Efficient formulation of the
  stochastic simulation algorithm for chemically reacting systems}, The journal
  of chemical physics, 121 (2004), pp.~4059--4067.

\bibitem{cappe2008adaptive}
{\sc O.~Capp{\'e}, R.~Douc, A.~Guillin, J.~Marin, and C.~Robert}, {\em Adaptive
  importance sampling in general mixture classes}, Statistics and Computing, 18
  (2008), pp.~447--459.

\bibitem{cappe2012population}
{\sc O.~Capp{\'e}, A.~Guillin, J.~Marin, and C.~Robert}, {\em {Population Monte
  Carlo}}, Journal of Computational and Graphical Statistics,  (2012).

\bibitem{chen2012monte}
{\sc M.~Chen, Q.~Shao, and J.~Ibrahim}, {\em {Monte Carlo methods in Bayesian
  computation}}, Springer Science \& Business Media, 2012.

\bibitem{chiavazzo2017intrinsic}
{\sc E.~Chiavazzo, R.~Covino, R.~Coifman, C.W. Gear, A.~Georgiou, G.~Hummer,
  and I.~Kevrekidis}, {\em Intrinsic map dynamics exploration for uncharted
  effective free-energy landscapes}, Proceedings of the National Academy of
  Sciences, 114 (2017), pp.~E5494--E5503.

\bibitem{constantine2014active}
{\sc P.~Constantine, E.~Dow, and Q.~Wang}, {\em Active subspace methods in
  theory and practice: applications to kriging surfaces}, SIAM Journal on
  Scientific Computing, 36 (2014), pp.~A1500--A1524.

\bibitem{cornuet2012adaptive}
{\sc J.~Cornuet, J.~Marin, A.~Mira, and C.~Robert}, {\em Adaptive multiple
  importance sampling}, Scandinavian Journal of Statistics, 39 (2012),
  pp.~798--812.

\bibitem{cotter2015parallel}
{\sc C.~Cotter, S.~Cotter, and P.~Russell}, {\em {Ensemble Transport Adaptive
  Importance Sampling}}, Submitted to SIAM Journal on Uncertainty
  Quantification,  (2019).

\bibitem{cotter2016constrained}
{\sc S.~Cotter}, {\em {Constrained approximation of effective generators for
  multiscale stochastic reaction networks and application to conditioned path
  sampling}}, Journal of Computational Physics, 323 (2016), pp.~265--282.

\bibitem{cotter2016error}
{\sc S.~Cotter and R.~Erban}, {\em Error analysis of diffusion approximation
  methods for multiscale systems in reaction kinetics}, SIAM Journal on
  Scientific Computing, 38 (2016), pp.~B144--B163.

\bibitem{cotter2013mcmc}
{\sc S.~Cotter, G.~Roberts, A.~Stuart, and D.~White}, {\em {MCMC methods for
  functions: modifying old algorithms to make them faster}}, Statistical
  Science, 28 (2013), pp.~424--446.

\bibitem{cotter2011constrained}
{\sc S.~Cotter, K.~Zygalakis, I.~Kevrekidis, and R.~Erban}, {\em A constrained
  approach to multiscale stochastic simulation of chemically reacting systems},
  The Journal of Chemical Physics, 135 (2011), p.~094102.

\bibitem{douc2007convergence}
{\sc R.~Douc, A.~Guillin, J.~Marin, and C.~Robert}, {\em Convergence of
  adaptive mixtures of importance sampling schemes}, The Annals of Statistics,
  35 (2007), pp.~420--448.

\bibitem{douc2007minimum}
{\sc R.~Douc, A.~Guillin, J-M Marin, and C.~Robert}, {\em Minimum variance
  importance sampling via population monte carlo}, ESAIM: Probability and
  Statistics, 11 (2007), pp.~427--447.

\bibitem{dsilva2016data}
{\sc C.~Dsilva, R.~Talmon, C.W. Gear, R.~Coifman, and I.~Kevrekidis}, {\em
  Data-driven reduction for a class of multiscale fast-slow stochastic
  dynamical systems}, SIAM Journal on Applied Dynamical Systems, 15 (2016),
  pp.~1327--1351.

\bibitem{duane1987hybrid}
{\sc S.~Duane, A.~Kennedy, B.~Pendleton, and D.~Roweth}, {\em {Hybrid Monte
  Carlo}}, Physics letters B, 195 (1987), pp.~216--222.

\bibitem{weinan2007nested}
{\sc W.~E, D.~Liu, and E.~Vanden-Eijnden}, {\em Nested stochastic simulation
  algorithms for chemical kinetic systems with multiple time scales}, Journal
  of computational physics, 221 (2007), pp.~158--180.

\bibitem{el2012bayesian}
{\sc T.~El~Moselhy and Y.~Marzouk}, {\em Bayesian inference with optimal maps},
  Journal of Computational Physics, 231 (2012), pp.~7815--7850.

\bibitem{erban2006gene}
{\sc R.~Erban, I.~Kevrekidis, D.~Adalsteinsson, and T.~Elston}, {\em Gene
  regulatory networks: A coarse-grained, equation-free approach to multiscale
  computation}, The Journal of chemical physics, 124 (2006), p.~084106.

\bibitem{flandrin2003stationarity}
{\sc P.~Flandrin, P.~Borgnat, and P-O Amblard}, {\em From stationarity to
  self-similarity, and back: Variations on the lamperti transformation}, in
  Processes with Long-Range Correlations, Springer, 2003, pp.~88--117.

\bibitem{gardiner2009stochastic}
{\sc C.~Gardiner}, {\em Stochastic methods}, vol.~4, springer Berlin, 2009.

\bibitem{gillespie1977exact}
{\sc D.~Gillespie}, {\em Exact stochastic simulation of coupled chemical
  reactions}, The Journal of Physical Chemistry, 81 (1977), pp.~2340--2361.

\bibitem{gillespie2007stochastic}
\leavevmode\vrule height 2pt depth -1.6pt width 23pt, {\em Stochastic
  simulation of chemical kinetics}, Annu. Rev. Phys. Chem., 58 (2007),
  pp.~35--55.

\bibitem{girolami2011riemann}
{\sc M.~Girolami and B.~Calderhead}, {\em {Riemann manifold langevin and
  hamiltonian monte carlo methods}}, Journal of the Royal Statistical Society:
  Series B (Statistical Methodology), 73 (2011), pp.~123--214.

\bibitem{guido2006bottom}
{\sc N.~Guido, X.~Wang, D.~Adalsteinsson, D.~McMillen, J.~Hasty, C.~Cantor,
  T.~Elston, and J.~Collins}, {\em A bottom-up approach to gene regulation},
  Nature, 439 (2006), p.~856.

\bibitem{gutenkunst2007universally}
{\sc R.~Gutenkunst, J.~Waterfall, F.~Casey, K.~Brown, C.~Myers, and J.~Sethna},
  {\em Universally sloppy parameter sensitivities in systems biology models},
  PLoS computational biology, 3 (2007), p.~e189.

\bibitem{jahnke2007solving}
{\sc T.~Jahnke and W.~Huisinga}, {\em Solving the chemical master equation for
  monomolecular reaction systems analytically}, Journal of Mathematical
  Biology, 54 (2007), pp.~1--26.

\bibitem{kaern2005stochasticity}
{\sc M.~Kaern, T.~Elston, W.~Blake, and J.~Collins}, {\em Stochasticity in gene
  expression: from theories to phenotypes}, Nature reviews. Genetics, 6 (2005),
  p.~451.

\bibitem{kass1995bayes}
{\sc R.~Kass and A.~Raftery}, {\em Bayes factors}, Journal of the American
  Statistical Association, 90 (1995), pp.~773--795.

\bibitem{kevrekidis2003equation}
{\sc I.~Kevrekidis, C.W. Gear, J.~Hyman, P.~Kevrekidis, O.~Runborg,
  C.~Theodoropoulos, et~al.}, {\em Equation-free, coarse-grained multiscale
  computation: Enabling mocroscopic simulators to perform system-level
  analysis}, Communications in Mathematical Sciences, 1 (2003), pp.~715--762.

\bibitem{martino2015adaptive}
{\sc L.~Martino, V.~Elvira, D.~Luengo, and J.~Corander}, {\em An adaptive
  population importance sampler: Learning from uncertainty}, IEEE Transactions
  on Signal Processing, 63 (2015), pp.~4422--4437.

\bibitem{martino2017layered}
\leavevmode\vrule height 2pt depth -1.6pt width 23pt, {\em Layered adaptive
  importance sampling}, Statistics and Computing, 27 (2017), pp.~599--623.

\bibitem{y.17:_handb_uncer_quant}
{\sc Y.~Marzouk, T.~Moselhy, M.~Parno, and A.~Spantini}, {\em Handbook of
  Uncertainty Quantification}, Springer, 2017, ch.~An introduction to sampling
  via measure transport.

\bibitem{parno2018transport}
{\sc M.~Parno and Y.~Marzouk}, {\em Transport map accelerated markov chain
  monte carlo}, SIAM/ASA Journal on Uncertainty Quantification, 6 (2018),
  pp.~645--682.

\bibitem{pinsker1960information}
{\sc M.~Pinsker}, {\em Information and information stability of random
  variables and processes},  (1960).

\bibitem{reich2013nonparametric}
{\sc S.~Reich}, {\em {A nonparametric ensemble transform method for Bayesian
  inference}}, SIAM Journal on Scientific Computing, 35 (2013),
  pp.~A2013--A2024.

\bibitem{russ2017parallel}
{\sc P.~Russell}, {\em Parallel MCMC methods and their application in inverse
  problems}, PhD thesis, School of Mathematics, University of Manchester, 2017.

\bibitem{singer2008non}
{\sc A.~Singer and R.~Coifman}, {\em Non-linear independent component analysis
  with diffusion maps}, Applied and Computational Harmonic Analysis, 25 (2008),
  pp.~226--239.

\bibitem{singer2009detecting}
{\sc A.~Singer, R.~Erban, I.~Kevrekidis, and R.~Coifman}, {\em Detecting
  intrinsic slow variables in stochastic dynamical systems by anisotropic
  diffusion maps}, Proceedings of the National Academy of Sciences, 106 (2009),
  pp.~16090--16095.

\bibitem{teh2016consistency}
{\sc Y.W. Teh, A.~Thiery, and S.~Vollmer}, {\em Consistency and fluctuations
  for stochastic gradient langevin dynamics}, The Journal of Machine Learning
  Research, 17 (2016), pp.~193--225.

\end{thebibliography}

\end{document}